\documentclass[12pt]{article}
\usepackage[english,activeacute]{babel}
\usepackage{natbib}
\usepackage{comment}
\usepackage{float}
\usepackage[hidelinks]{hyperref}
\usepackage{mathrsfs}
\usepackage{enumitem}
\usepackage[font={small,it}]{caption}
\usepackage{amsmath,amsfonts,amsthm,amssymb}
\usepackage{bm,rotating,multirow,dsfont,graphicx}
\usepackage[usenames, dvipsnames]{color}
\usepackage{url}
\usepackage{multicol}
\usepackage{multirow}
\usepackage[T1]{fontenc}
\usepackage{flafter}
\usepackage{appendix}
\usepackage{subfigure}
\usepackage{xcolor}
\usepackage{soul}
\makeatletter
\def\hlinewd#1{%
	\noalign{\ifnum0=`}\fi\hrule \@height #1 %
	\futurelet\reserved@a\@xhline}
\makeatother
\def\spacingset#1{\renewcommand{\baselinestretch}{#1}\small\normalsize}\spacingset{1}
\def\@roman#1{\romannumeral #1}
\begin{document}

\title{Identifying Hierarchical Structures in Network Data}

\date{}

\author{
    Pedro Regueiro, Nubank, Mexico \\
    Abel Rodríguez, University of Washington, US \\
    Juan Sosa, Universidad Nacional de Colombia, Colombia\footnote{Corresponding author: jcsosam@unal.edu.co.}
}

\maketitle

\begin{abstract} 
In this paper, we introduce a hierarchical extension of the stochastic blockmodel to identify multilevel community structures in networks. We also present a Markov chain Monte Carlo (MCMC) and a variational Bayes algorithm to fit the model and obtain approximate posterior inference. Through simulated and real datasets, we demonstrate that the model successfully identifies communities and supercommunities when they exist in the data. Additionally, we observe that the model returns a single supercommunity when there is no evidence of multilevel community structure.
As expected in the case of the single-level stochastic blockmodel, we observe that the MCMC algorithm consistently outperforms its variational Bayes counterpart. Therefore, we recommend using MCMC whenever the network size allows for computational feasibility.

\end{abstract}


\spacingset{1.1} % DON'T change the spacing!

\section{Introduction}\label{intro}

Network science is an interdisciplinary field of study that borrows from advances in mathematics, physics, statistics, sociology, computer science and machine learning, among others. This has lead to a diverse set of tools and techniques that improve our understanding of complex systems. For example, models such as the \emph{preferential attachment} of \citep{Price76,Barabasi&Albert99} and the \emph{small world model} of \cite{Watts&Strogatz98} help us understand the process of network formation, whereas works like \cite{Grassberger83} and \cite{Pastor-Satorras&Vespignani01} shed light into how processes evolve in a network. A good overview of the development of the field can be found in \cite{Newman03} or \cite{Newman10} and, in particular, an overview of the statistically oriented literature can be found in \cite{Goldenbergetal10}.

Perhaps the best well studied problem within network analysis is that of \emph{community detection}, which refers to the splitting a graph into \emph{communities} sometimes also referred to as \emph{clusters}. As accounted in \citep{Fortunato10, Newman04, Schaeffer07, Porteretal09}, many different solutions to this problem have been proposed in the literature.  Among the most successful approaches are those based on the ideas of \emph{hierarchical clustering}, \emph{betweenness}, and \emph{modularity}. However, the majority of methods developed in this area are deterministic algorithms, which do not provide a measure of the uncertainty associated with the solutions they generate. Also, a drawback from this literature is the fact hat these algorithms are usually developed to recover \emph{assortative} structures (groups of vertices with high density of connections within each group, but few interactions across them).

Furthermore, a feature that is commonly observed in network data is the hierarchical structure of communities. That is, nested arrangements in which vertices group to form communities and, in turn, communities group into so-called \emph{supercommunities} or \emph{metacommunities}. 
That is why here we introduce a hierarchical extension of the \emph{stochastic blockmodel} \citep{Hollandetal83} that is able to capture the multilevel structure of communities. 
Our work is closest to that of \cite{Hoetal12}, though we use a fundamentally different approach to introduce the hierarchy in the community parameters. 
Among the most appealing features of stochastic block models is their capacity to simultaneously recover both assortative and disassortative communities, along with their ability to quantify the uncertainty associated with the resulting partition structure.

The reminder of this paper is organized as follows: Section \ref{bb} reviews the basic ideas behind stochastic blockmodels, introduces a hierarchical extension that allows to recover multilevel community structures, and discusses choices for prior distribution specification. Section \ref{psml} describes a posterior sampling scheme for this model using \textit{Markov chain Monte Carlo} (MCMC) techniques. As discussed there, the heavy computational requirements of this approach makes MCMC impractical or even infeasible for moderately large networks; for this reason, we also introduce in this section a \emph{variational} algorithm that allows to obtain approximate posterior inference. Section \ref{ilust} presents various illustrations utilizing both synthetic and real datasets, comparing the results obtained from the two methods employed for posterior inference. Finally, \ref{disc} summarizes our findings, discusses its implications, and presents directions for further research.

\section{Model}
\label{bb}

\subsection{Stochastic blockmodels}
\label{slbb}
The stochastic blockmodel \citep{Hollandetal83} is a simple, yet very flexible model that allows to represent different kind of interactions among different types of agents in a complex system. Here, we concentrate on the case of \emph{simple}, \emph{unweighted} and \emph{undirected} networks, which can be characterized in terms of their \emph{adjacency matrix} or \emph{sociomatrix}, $Y\in\mathbb{R}^{I \times I},$ whose entries are given by  
\begin{equation*}
y_{i,j}=\begin{cases} 1, \quad &\text{if there is an edge between vertices $j$ and $i$;}\\ 0, \quad &\text{otherwise,}\end{cases}
\end{equation*}
where $I$ represent the number of vertices in the network. Furthermore, because of the symmetry constraint imposed by the assumption of undirectedness, we focus on the strictly upper triangular part of the adjacency matrix 
\begin{equation*}
\mathcal{Y}=\{y_{i,j}: i,j\in\mathbb{N},\; 1\leq i<j\leq I\}.
\end{equation*}

The binary nature of interactions naturally suggests modeling them through a Bernoulli distribution
\begin{equation}\label{lm1}
y_{i,j}\mid \lambda_{i,j} \sim \mathsf{Ber}(\lambda_{i,j}), \qquad 1\leq i<j\leq I.
\end{equation}
Now, the basic idea behind the stochastic blockmodel is that the network can be partitioned into $K\leq I$ communities, where two vertices are in the same community only if they have equal interaction probabilities across the network. Formally, 
\begin{equation*}
\lambda_{i,j}=g(\theta_{\xi_i,\xi_j}),
\end{equation*}
where the block indicators $\xi_1,\ldots,\xi_I$ take their values in the set $\{1,\ldots,K\},$  the elements of the set $\{\theta_{k,l}\}_{k,l=1}^K$ are usually referred to as the community parameters, and $g(\cdot)$ is an appropriate link function. Notice that, again, because of symmetry, attention can be constraint to the set 
\begin{equation*}
\Theta=\{\theta_{k,l}: k,l\in\mathbb{N},\; 1\leq k\leq l \leq K\}.
\end{equation*}

Assuming conditional independence in the interactions both within and across actors, the likelihood can be expressed as  
\begin{equation}\label{likeasprod}
p(\mathcal{Y}\mid \Theta,\boldsymbol{\xi})=\prod_{i=1}^{I-1}\prod_{j=i+1}^I p\left(y_{i,j}\mid \theta_{\phi\left(\xi_i,\xi_j\right)}\right),
\end{equation}
where $\phi:\mathbb{R}^2\rightarrow\mathbb{R}^2$ is function given by $\phi(u,v)=(\min\{u,v\},\max\{u,v\})$, which enforces the symmetry conditions. In turn, \eqref{likeasprod} implies that 
\begin{equation}\label{likeprodgen}
p(\mathcal{Y}\mid \Theta,\boldsymbol{\xi})=\prod_{k=1}^K\prod_{l=k}^K\left\{g(\theta_{k,l})\right\}^{s_{k,l}}\left\{1-g(\theta_{k,l})\right\}^{n_{k,l}-s_{k,l}},
\end{equation}
with $s_{k,l}=\sum_{\mathcal{S}_{k,l}}y_{i,j}$ and $n_{k,l}=\sum_{\mathcal{S}_{k,l}}1,$ and the sum is taken over the set 
\begin{equation*}
\mathcal{S}_{k,l}=\left\{(i,j):\; i,j\in\mathbb{N},\;i<j,\; (k,l)=\phi\left(\xi_i,\xi_j\right)\right\}.
\end{equation*}

Regarding the maximum number of communities $K,$ a popular alternative consists in following a nonparametric approach as in the \emph{infinite relational model} of \cite{Kempetal06}. This model allows for the effective number of communities in the network $K^\star\leq K$ to be learned from the data, being able to take any integer value between 1 and $I.$ Specifically, $\boldsymbol{\xi}$ is assumed to follow a \emph{Chinese restaurant process} (CRP) prior, which implies that its distribution is given by Ewens sampling formula \citep{Ewens72}, that is 
\begin{equation*}
p(\xi_1,\ldots,\xi_I)=\frac{\Gamma(\alpha)\alpha^{K^\star}}{\Gamma(\alpha+I)}\prod_{k=1}^{K^\star}\Gamma(n_k).
\end{equation*}
Alternatively, $K$ can be fixed trying to overestimate the number of communities in the network, thus leading to a finite mixture model as in \cite{Nowicki&Snijders01}.  Here, for simplicity, we adopt this later approach. Thus, we assume that the entries of $\boldsymbol{\xi}$ are exchangeable and follow a Categorical distribution in $\{1,\ldots,K\}$ 
\begin{equation}\label{pxim1}
\textsf{Pr}(\xi_i=k\mid w_k)=w_k,\qquad i=1,\ldots,I,
\end{equation}
with the weights vector $\boldsymbol{w}$ satisfying 
\begin{equation}
\boldsymbol{w} \sim \textsf{Dir}\left(\boldsymbol{\alpha_{w}}\right). 
\end{equation}
As discussed in \cite{Ishwaran&Zarepour00} and \cite{Neal00}, if the parameter vector is set as $\boldsymbol{\alpha_{w}}=\left(\frac{\alpha}{K},\ldots,\frac{\alpha}{K}\right),$ then as $K\rightarrow\infty$, the model approximates the infinite relational model.

For the prior specification, we use a logit structure under a Gaussian prior for the elements of $\Theta.$ Specifically, if $g$ is taken to be the canonical link, that is, $\theta_{\xi_i,\xi_j}=\log\left(\frac{\lambda_{i,j}}{1-\lambda_{i,j}}\right),$ for each element in $\mathcal{Y}$, we have that 
\begin{equation*}
p(y_{i,j}\mid \theta_{\xi_i,\xi_j})=\frac{(\exp\{\theta_{\phi\left(\xi_i,\xi_j\right)}\})^{y_{i,j}}}{1+\exp\{\theta_{\phi\left(\xi_i,\xi_j\right)}\}},
\end{equation*}
which reduces the likelihood to  
\begin{equation}\label{likeprod}
p(\mathcal{Y}\mid \Theta,\boldsymbol{\xi})=\prod_{k=1}^K\prod_{l=k}^K\frac{(\exp\{\theta_{k,l}\})^{s_{k,l}}}{(1+\exp\{\theta_{k,l}\})^{n_{k,l}}}.
\end{equation}

Now, the model can be completed with hyperpriors. Specifically, we assume the community parameters conditionally independent from a common Gaussian prior
\begin{equation}\label{petam1}
\theta_{k,l}\mid \mu,\sigma^2\sim \textsf{N}(\mu,\sigma^2).
\end{equation}
In this case, $\mu$ affects the overall density of the network, while $\sigma^2$ control the variability among the propensity of interaction between different clusters. Thus, setting $\mu=0$ centers the interaction probabilities at $\frac{1}{2},$ while, considering the transformation, choosing $\sigma^2=1$ leaves approximately $95\%$ of the mass in $(0.12,0.88)$ a priori, for all $\lambda_{i,j}.$ If instead a hyperprior $\pi(\mu,\sigma^2)$ is to be placed in these parameters, one computationally convenient option is choosing conditionally conjugate distributions
\begin{equation}
\mu \sim \textsf{N}(\mu_\mu,\sigma^2_\mu) \quad  \text{and} \quad \sigma^2  \sim \textsf{IG}(\alpha_{\sigma},\beta_{\sigma}).
\end{equation}

The concentration parameter $\alpha$ controls the number of occupied communities in the network $K^\star.$ In the limit case of the infinite relational model, from \cite{Antoniak74}, it is known that the distribution satisfies     
\begin{equation*}
\textsf{Pr}(K^\star=k\mid\alpha)=S(I,k)\,\alpha^k\frac{\Gamma(\alpha)}{\Gamma(\alpha+I)},
\end{equation*}
where $S$ represents the unsigned Sterling numbers of the first kind. Thus, 
 \begin{equation*}
\textsf{E}[K^\star\mid\alpha]=\alpha\left[\Psi(\alpha+I)-\Psi(\alpha)\right]\approx \alpha \log\left(\frac{\alpha+I}{\alpha}\right)
\end{equation*}
and 
 \begin{equation*}
\textsf{Var}[K^\star\mid\alpha]=\alpha\left[\Psi(\alpha+I)-\Psi(\alpha)\right]+\alpha^2\left[\Psi^\prime(\alpha+I)-\Psi^\prime(\alpha)\right]\approx \alpha \log\left(\frac{\alpha+I}{\alpha}\right),
\end{equation*}
with the first order approximations valid for large $I.$

\begin{figure}[!htb]
\centering
\includegraphics[width=.6\textwidth]{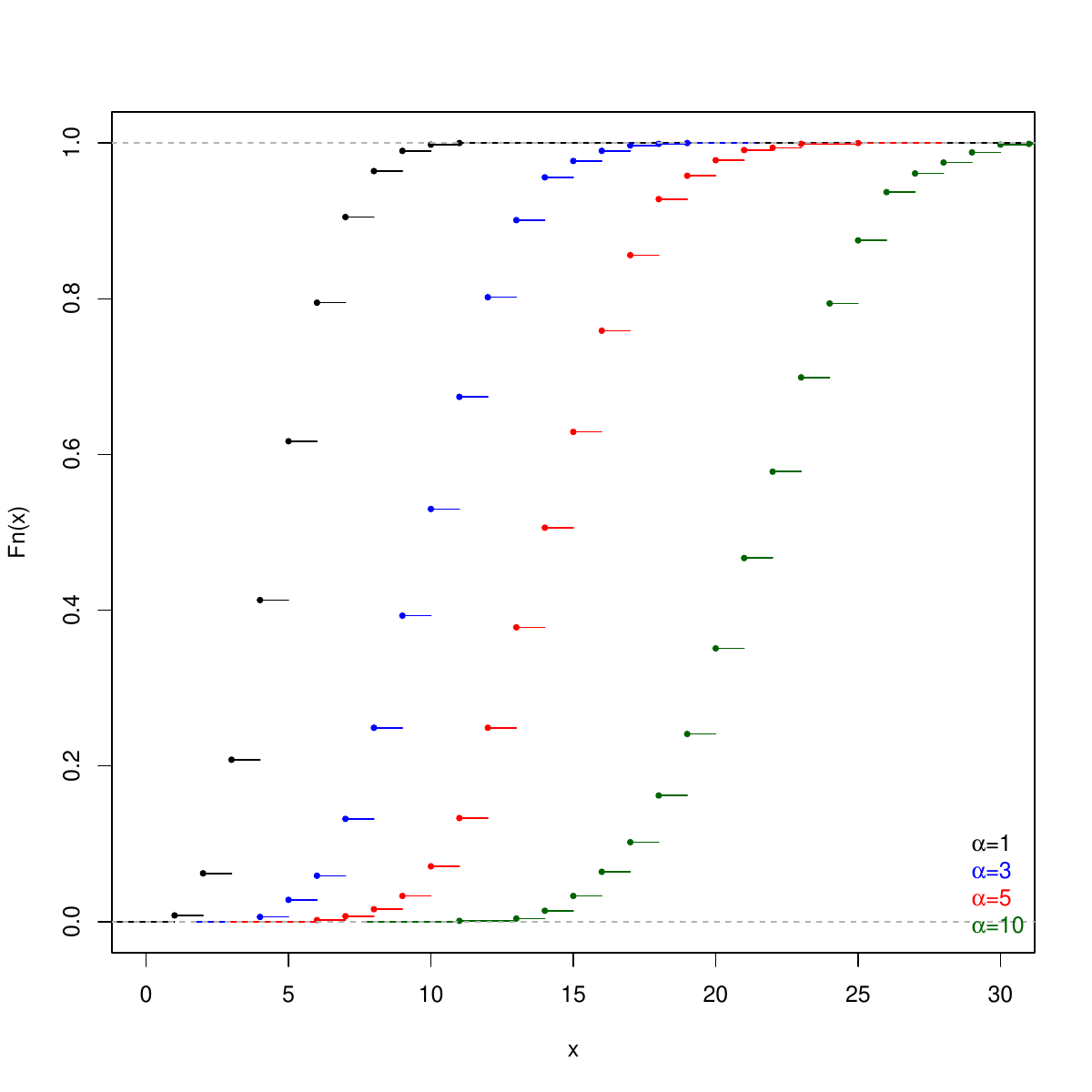} 
\caption{Comulative distribution function of the effective number of communities $K^\star$ implied by the prior for four different values of $\alpha,$ $\alpha=1$ (black), $\alpha=3$ (blue), $\alpha=5$ (red), and $\alpha=10$ (green) when $K=I=100$.}
\label{alphaprior}
\end{figure}

Here, we explore the effect of $\alpha$ via simulation. As an example, Figure \ref{alphaprior} shows the empirical cumulative distribution function of $K^\star$ for four different values of $\alpha$ in the case where $I=K=100.$ This figure shows that, in this case, the behavior expected from the nonparametric model is also observed for a finite $K$; specifically, the expected number of clusters increases with $\alpha$, approximately following the order of $\alpha\log(I/\alpha)$. If alternatively, $\alpha$ is to be learned from the data, a common choice of hyperprior is
\begin{equation}
\alpha\sim\textsf{G}(\alpha_\alpha,\beta_\alpha),
\end{equation}
with the Gamma distribution parameterized in terms of shape and rate.

\subsection{A multilevel stochastic blockmodel}
\label{mlbb}

A desirable property of the stochastic blockmodel is that it can easily be generalized in a hierarchical fashion and, thus, it naturally lends itself for the problem at hand: The discovery of multilevel structures in networks. 
Specifically, we assume that the community parameters come from a mean mixture of Gaussians 
\begin{equation}\label{pthetam1ml}
\theta_{k,l}\,|\,\eta_{\zeta_k,\zeta_l},\sigma^2\sim \textsf{N}(\eta_{\phi(\zeta_k,\zeta_l)},\sigma^2),
\end{equation}
where the location parameters $\eta_{r,s}$ are assumed conditionally independent from a common Normal distribution 
\begin{equation}\label{petam1ml}
\eta_{r,s}\,|\,\mu,\sigma^2\sim \textsf{N}(\mu,\tau^2),
\end{equation}
and, once again, symmetry restrictions allow to consider only the subset
\begin{equation*}
H=\{\eta_{r,s}: r,s\in\mathbb{N},\, 1\leq r\leq s \leq R\}.
\end{equation*}
As in the single-level model, \(\sigma^2\) regulates the variability in the propensity for interactions between communities within a supercommunity, and \(\mu\) governs the overall network density. Meanwhile, at the second level, \(\tau^2\) controls the dispersion of the community parameter means.

The choice of prior for the supercommunity indicators $\boldsymbol{\zeta}$ mimics the structure of the first level indicators, taking a Categorical distribution in the set $\{1,\ldots,R\}$, with 
\begin{equation}\label{pxim1ml}
\textsf{Pr}(\zeta_k=r\mid v_r)=v_r,\qquad k=1,\ldots,K,
\end{equation}
with the weights vector $\boldsymbol{v}$ such that
\begin{equation}\label{hpm2ml}
\boldsymbol{v} \sim \textsf{Dir}\left(\boldsymbol{\alpha_{v}}\right),
\end{equation}
where $\boldsymbol{\alpha_{v}}=\left(\frac{\beta}{R},\ldots,\frac{\beta}{R}\right).$
Analogous to the single level model we have that, as $K\rightarrow\infty,$
\begin{equation*}
\textsf{Pr}(K^\star=k\mid\alpha)=S(I,k)\,\alpha^k\frac{\Gamma(\alpha)}{\Gamma(\alpha+I)},
\end{equation*}
and further, when $R\;\rightarrow\;\infty,$
\begin{equation*}
\textsf{Pr}(R^\star=r\mid K^\star,\beta)=S(K^\star,r)\,\beta^r\frac{\Gamma(\beta)}{\Gamma(\beta+K^\star)}.
\end{equation*}

Then, conditionally, the expected number of supercommunities can be approximated as
\begin{equation*}
\textsf{E}[R^\star\mid\beta,K^\star]\approx \beta\log\left(\frac{\beta+K^\star}{\beta}\right),
\end{equation*}
and the expected number of supercommunites is found to satisfy
\begin{equation*}
\textsf{E}[R^\star\mid\alpha,\beta]=\textsf{E}[\textsf{E}[R^\star\mid\beta,K^\star]\mid\alpha]\approx \beta\log\left(\frac{\beta+\alpha\log\left(\frac{\alpha+I}{\alpha}\right)}{\beta}\right)-\frac{\beta\alpha\log\left(\frac{\alpha+I}{\alpha}\right)}{2\left(\beta+\alpha\log\left(\frac{\alpha+I}{\alpha}\right)\right)^2}.
\end{equation*}
The details of this derivation can be found in Appendix \ref{AENC}.

Figure \ref{abprior} shows the impact of the hyperparameters \(\alpha\) and \(\beta\) on the number of occupied communities and supercommunities implied by the prior. This revision improves clarity and flow while preserving the original meaning. We see that, as it is usual, larger values of the concentration parameters favor a larger number of components at both levels. We note also that the standard stochastic blockmodel can then be recover by either setting $R=1$ or letting $\beta\rightarrow 0.$

\begin{figure}[!htb]
\centering
\begin{tabular}{@{}cc@{}}
    \includegraphics[width=.47\textwidth]{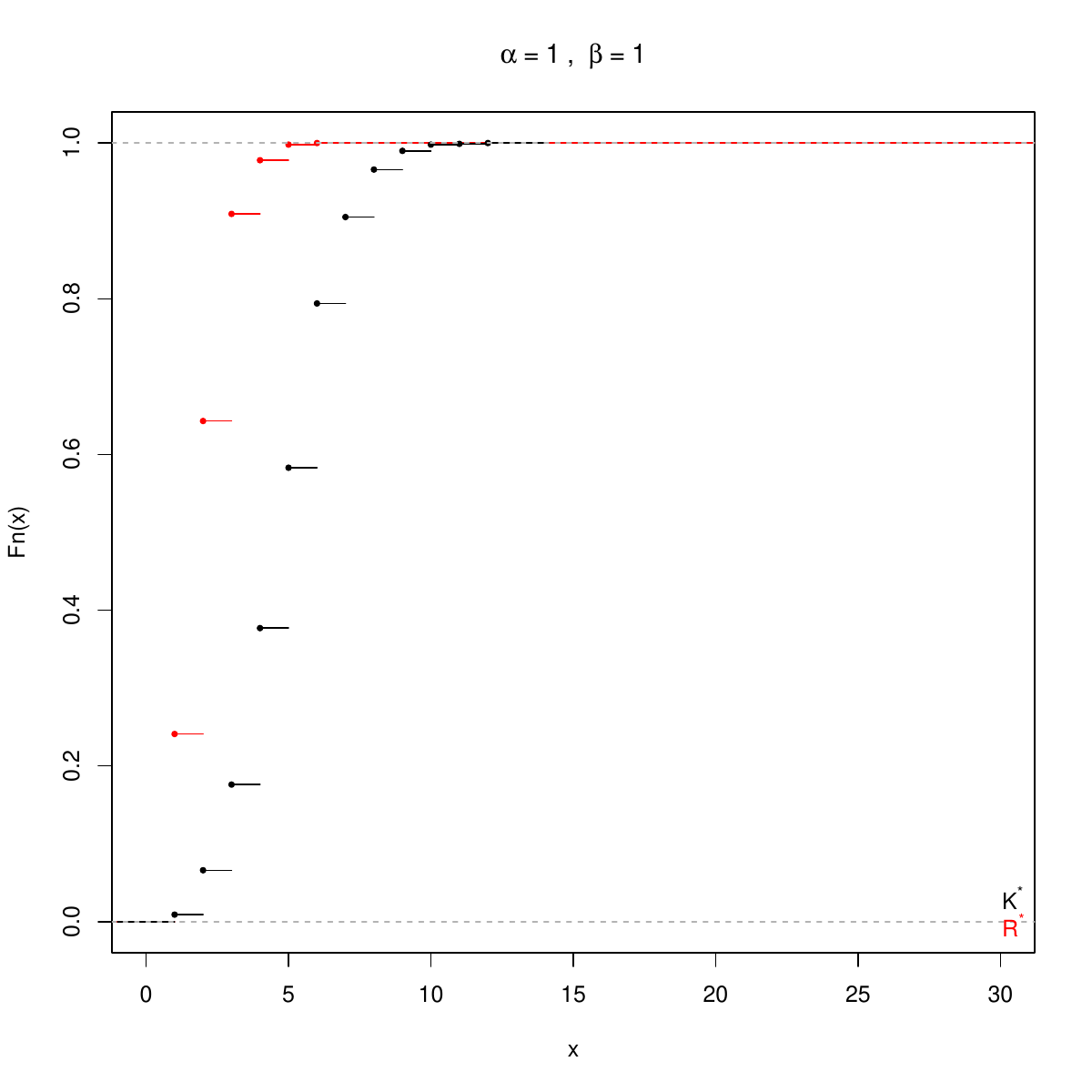} &
    \includegraphics[width=.47\textwidth]{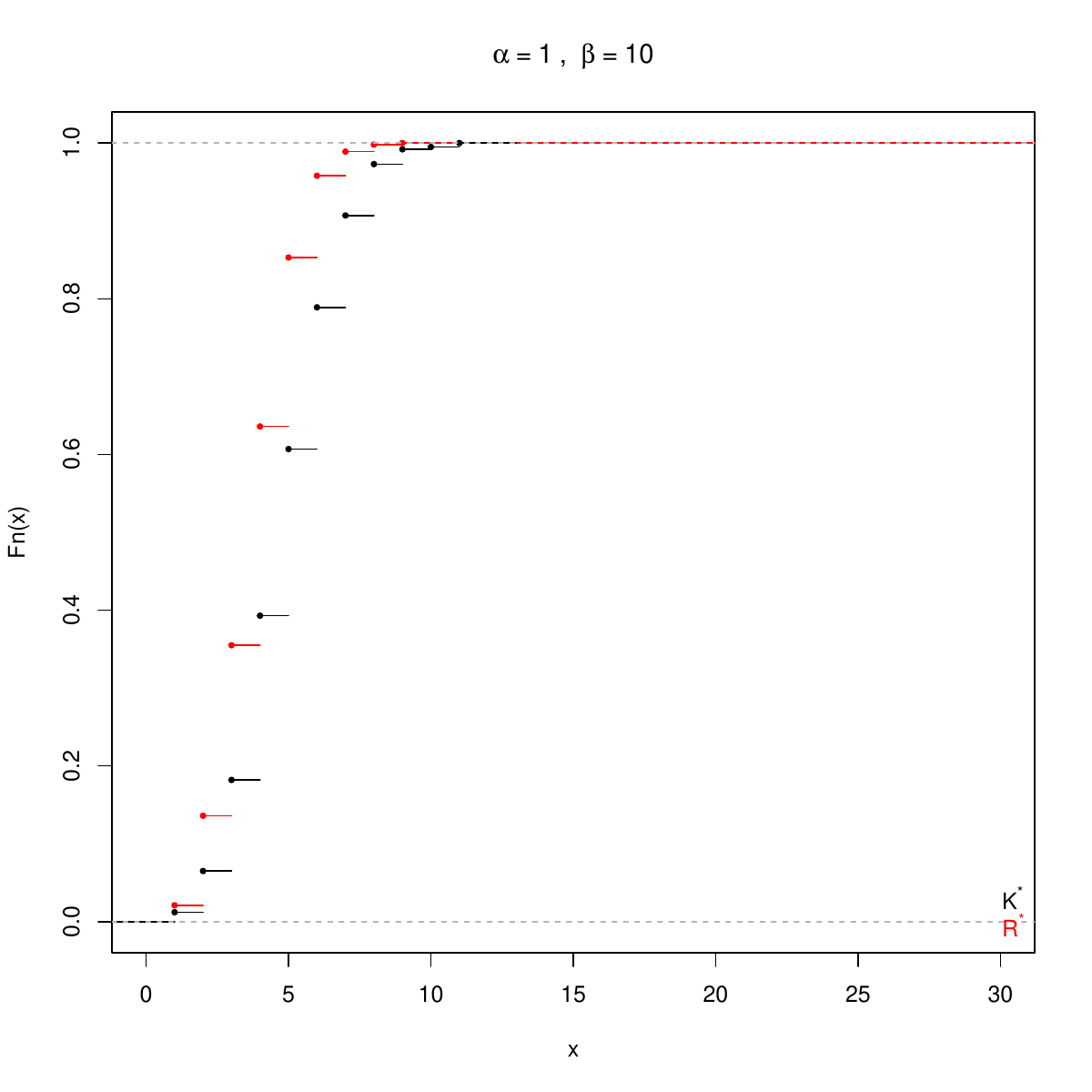}  \\  
    \includegraphics[width=.47\textwidth]{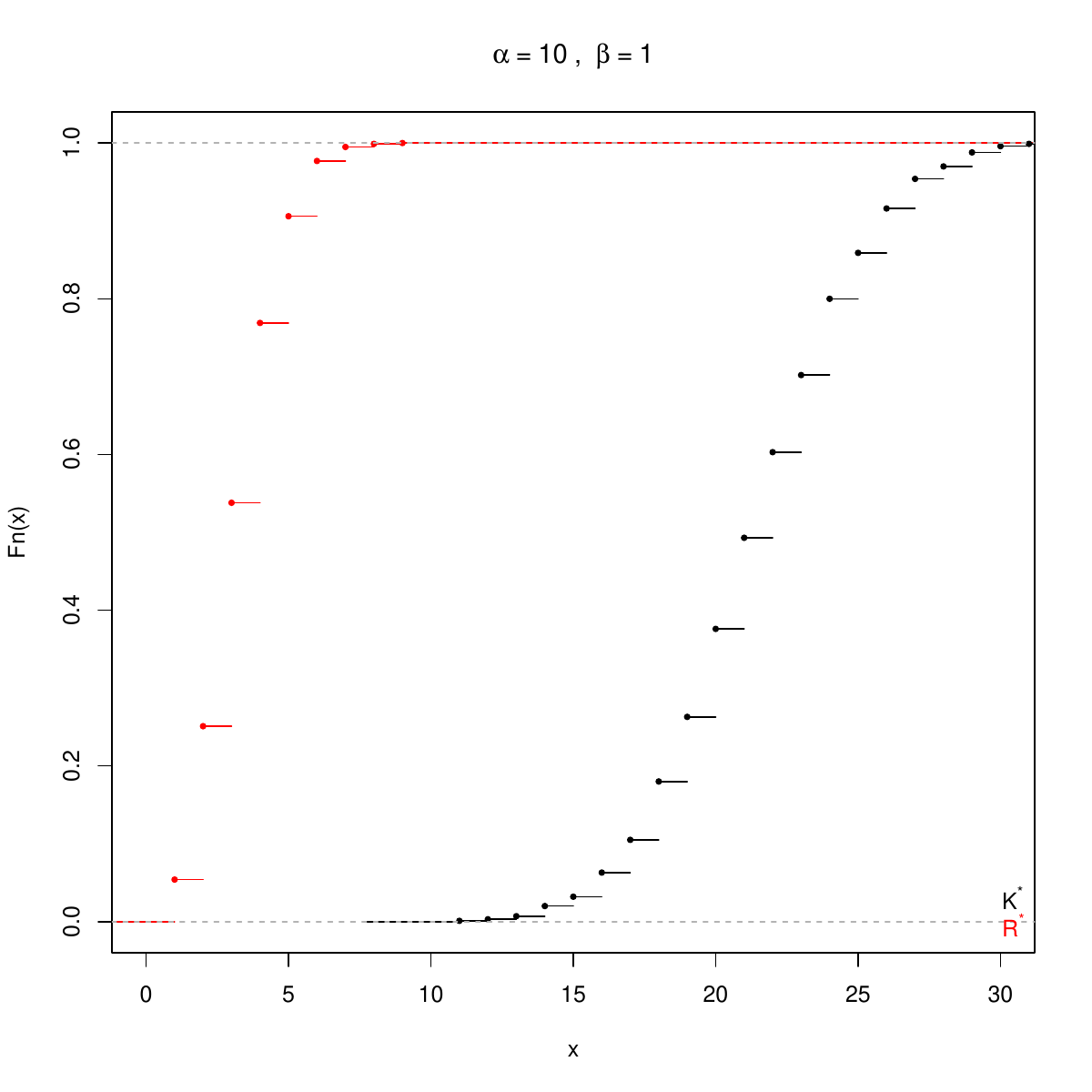} &
    \includegraphics[width=.47\textwidth]{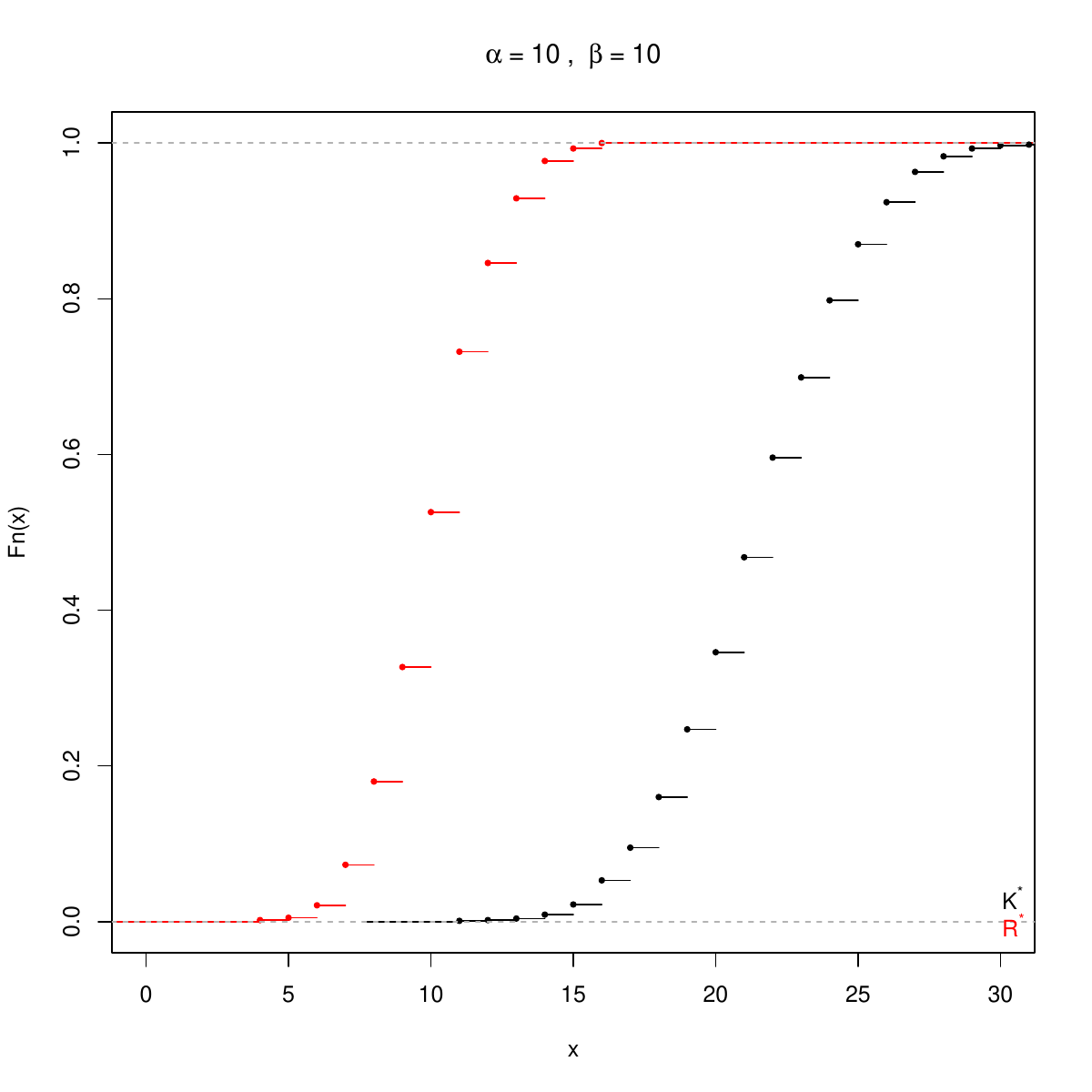}  
\end{tabular}
\caption{Prior cumulative distribution function corresponding to the effective number of communities $K^\star$ and supercommunities $R^\star$ under four different scenarios for the hyperparameters $\alpha$ and $\beta.$}
\label{abprior}
\end{figure}

Finally, all hyperparameters are assumed independent a priori,
\begin{equation*}
p(\mu,\sigma^2,\tau^2,\alpha,\beta)=p(\mu)\,p(\sigma^2)\,p(\tau^2)\,p(\alpha)\,p(\beta),
\end{equation*}
with conditionally conjugate hyperpriors in the community parameter's side 
\begin{equation}\label{hpm1ml}
\mu \sim \textsf{N}(\mu_0,\sigma^2_\mu), \quad \tau^2  \sim \textsf{IG}(\alpha_{\tau},\beta_{\tau}) \quad  \text{and} \quad \sigma^2  \sim \textsf{IG}(\alpha_{\sigma},\beta_{\sigma}),
\end{equation}
and Gamma hyperpriors for the concentration parameters
\begin{equation}
\alpha\sim\textsf{G}(\alpha_\alpha,\beta_\alpha) \quad  \text{and} \quad \beta\sim\textsf{G}(\alpha_\beta,\beta_\beta).
\end{equation}

\section{Posterior inference}\label{psml}

\subsection{MCMC sampling}

Regardless of the choice of prior distributions, the model described above does not yield closed-form posteriors, and, thus, some form of approximation is required for inferential purposes. In this section, we derive a Markov chain Monte Carlo (MCMC) algorithm to generate samples from the joint posterior distribution of the model \citep{Geman&Geman84, Gelfand&Smith90}.

As a first step, notice that the form of the likelihood already suggests that the full conditionals for $\Theta$ are not members of a standard family of distributions. However, following \cite{Polsonetal13}, 
\begin{equation*}
\frac{(\exp\{\theta_{k,l}\})^{s_{k,l}}}{(1+\exp\{\theta_{k,l}\})^{n_{k,l}}} =\exp\left\{\left(s_{k,l}-\frac{n_{k,l}}{2}\right)\theta_{k,l}\right\}\textsf{E}\left[ \exp\left\{-\frac{\theta_{k,l}^2}{2}\gamma_{k,l}\right\}\right]
\end{equation*}
where $\gamma_{k,l} \sim \textsf{PG}(n_{k,l},0)$ is a Polya-Gamma random variable. Thus, augmenting the parameter space with a matrix of a priori independent Polya-Gamma random variables $\Gamma=[\gamma_{k,l}]$, the likelihood can be expressed as 
\begin{equation}\label{likePS}
p(\mathcal{Y}\mid \Theta,\boldsymbol{\xi},\Gamma)\propto \exp\left\{\sum_{k=1}^K\sum_{l=k}^K\left(s_{k,l}-\frac{n_{k,l}}{2}\right)\theta_{k,l}\right\}\prod_{k=1}^K\prod_{l=k}^K \exp\left\{-\frac{\theta_{k,l}^2}{2}\gamma_{k,l}\right\}.
\end{equation}
Denoting the set of all parameters in the model $\Upsilon=\left\{\Theta,\boldsymbol{\xi},H,\boldsymbol{\zeta},\mu,\sigma^2,\tau^2,\boldsymbol{w},\alpha,\boldsymbol{v},\beta\right\},$ the augmented joint posterior satisfies
\begin{multline} \label{postml}
p(\Upsilon, \Gamma \mid\mathcal{Y}) \propto  \exp\left\{\sum_{k=1}^K\sum_{l=k}^K\left(s_{k,l}-\frac{n_{k,l}}{2}\right)\theta_{k,l}\right\}\prod_{k=1}^K\prod_{l=k}^K \textsf{E}\left[\exp\left\{-\frac{\theta_{k,l}^2}{2}\gamma_{k,l}\right\}\right]\\
 (\sigma^2)^{-\left(\frac{1}{4}K(K+1)+\alpha_\sigma+1\right)} \exp\left\{-\frac{1}{2\sigma^2}\sum_{k=1}^K\sum_{l=k}^K(\theta_{k,l}-\eta_{\phi\left(\zeta_k,\zeta_l\right)})^2-\frac{\beta_\sigma}{\sigma^2}\right\}\exp\left\{-\frac{1}{2\sigma^2_\mu}(\mu-\mu_0)^2\right\}\\
 (\tau^2)^{-\left(\frac{1}{4}R(R+1)+\alpha_\tau+1\right)} \exp\left\{-\frac{1}{2\tau^2}\sum_{r=1}^R\sum_{s=r}^R(\eta_{r,s}-\mu)^2-\frac{\beta_\tau}{\tau^2}\right\} \prod_{k=1}^K w_k^{\frac{\alpha}{K}+n_k-1} \prod_{r=1}^R v_r^{\frac{\beta}{R}+m_r-1} \\
 \frac{\Gamma(\alpha)}{\left[\Gamma\left(\frac{\alpha}{K}\right)\right]^K} \alpha^{\alpha_\alpha-1} \exp\left\{-\beta_\alpha\alpha\right\} \frac{\Gamma(\beta)}{\left[\Gamma\left(\frac{\beta}{R}\right)\right]^R} \beta^{\alpha_\beta-1} \exp\left\{-\beta_\beta\beta\right\}\, p(\Gamma) 
\end{multline}
where $m_r=\sum_{\mathcal{T}_r}1$, with $\mathcal{T}_r=\{k:\;\zeta_k=r\}$, for all $r\in\{1,\ldots,R\}.$  
From \eqref{postml}, we derive an MCMC algorithm that enables sampling-based approximate inference for the model. The main ideas behind this algorithm are summarized in what follows, while the details can be found in Appendix \ref{AMCMCmult}.

First, from the form of the augmented likelihood \eqref{likePS}, it can be anticipated that the community parameters are conditionally conjugate given the auxiliary variables. Thus, the elements of $\Theta$ are sample from their corresponding Gaussian full conditional distribution. Importantly, the full conditional distribution for the auxiliary parameters remains in the Polya-Gamma family and can, therefore, be sampled as described in \cite{Polsonetal13}.

The communities and the supercommunities indicators are sampled from their Categorical full conditional distributions. At the supercommunity level the clustering probabilities are affected from the data trough the  term \[\exp\left\{-\frac{1}{2\sigma^2}\sum_{k=1}^K\sum_{l=k}^K(\theta_{k,l}-\eta_{\phi\left(\zeta_k,\zeta_l\right)})^2\right\},\] which plays a role equivalent to that of the likelihood in the community level.

The full conditional for $\eta_{r,s}$ is also Gaussian with mean and precision parameters that can be expressed as linear combinations of the prior mean $\mu$ and the proportion of observed interactions among the vertices in supercommunities $r$ and $s.$ In turn, if conditionally conjugate priors are assumed, the full conditional distribution for $\mu$, $\sigma^2$, and $\tau^2$ are straightforward (Gaussian for $\mu$ and Inverse Gamma for $\sigma^2$ and $\tau^2$). Finally, the concentration parameters $\alpha$ and $\beta$ can be sampled following ideas provided in \cite{Escobar&West95}.

\subsection{Variational approximations}

The Gibbs sampler described in the previous section is a standard tool in Bayesian inference. This algorithm is theoretically guaranteed to converge eventually to the posterior distribution $p(\Upsilon\mid\mathcal{Y}),$ which enables us to control the accuracy of the approximation by adjusting the number of generated samples. In practice, however, MCMC methods can be computationally intensive. Furthermore, a large number of iterations can be required to guarantee convergence of the algorithm, because of the multimodality of the posterior distribution. Thus, posterior sampling approaches can be impractical or infeasible, even for moderately large networks. For this reason, in the this section we explore a {variational Bayes} algorithm, an alternative technique that aims to functionally approximate the posterior distribution.

The main idea can be briefly summarized as follows. The purpose is to approximate $p(\Upsilon\mid\mathcal{Y})$ by another function $q(\Upsilon)$ that is restricted to be a member of a certain family of functions. To this end, it is possible to define a measure of divergence and use calculus of variation techniques to find that $q$ in such family that minimizes this measure. Applying this technique to cases where $p$ is chosen as a posterior distribution and the Kullback–Leibler (KL) divergence is used can be traced back to works such as \cite{Sauletal96} and \cite{Jordanetal99}. In this case the problem becomes 
\begin{equation}\label{vbproblem}
\underset{q}{\min} \int q(\Upsilon) \log \frac{q(\Upsilon)}{p(\Upsilon\mid\mathcal{Y})} \,\textsf{d}\Upsilon.
\end{equation}

Now, it is easy shown that  
\begin{equation*}
\int q(\Upsilon) \log \frac{q(\Upsilon)}{p(\Upsilon\mid\mathcal{Y})} \,\textsf{d}\Upsilon=\log p(\mathcal{Y}) -  \int q(\Upsilon) \log \frac{p(\Upsilon,\mathcal{Y})}{q(\Upsilon)} \,\textsf{d}\Upsilon
\end{equation*} 
and therefore, minimizing KL$[q\mid\mid p]$ is equivalent to maximizing $\textsf{E}_{q(\Upsilon)}\left[\log \frac{p(\Upsilon,\mathcal{Y})}{q(\Upsilon)}\right]$ which is known as the \emph{evidence lower bound} (ELBO). Note further that the ELBO can be expressed as 
\begin{equation*}
F(q,\mathcal{Y})=\textsf{E}_{q(\Upsilon)}[\log p(\mathcal{Y},\Upsilon)]+H\left[q(\Upsilon)\right]
\end{equation*}
where $H$ denotes the Shannon entropy. Furthermore, if $q$ is assumed to satisfy the mean field assumption
\begin{equation*}
q(\Upsilon)=\prod_{i} q_i(\Upsilon_{i})
\end{equation*}
where $q_i(\Upsilon_{i})$ are the marginal variational densities, the solution of this problem satisfies
\begin{equation}\label{vbgensolution}
\log q^\star_i(\Upsilon_{i})\propto\textsf{E}_{q(\boldsymbol{\Upsilon_{-i}})}\left[ \log p(\Upsilon,\mathcal{Y}) \right] 
\end{equation}
which leads to a coordinate optimization algorithm \citep{Dempsteretal77}. From equation \eqref{vbgensolution} we can proceed to obtain the optimal variational distributions for the model parameters.

Similar to the corresponding full conditional, the variational distribution of the elements of $\Theta$ does not belong to a standard family of distributions. In order to overcome this issue, a first alternative would be to introduce Polya-Gamma auxiliary variables in the form of \eqref{likePS}; this conduces to a Gaussian update for the elements of $\Theta,$ but translates into non standard updates for the auxiliary variables. In turn, these updates could be handled using, for example, ideas from non-conjugate variational message passing \citep{Knowles&Minka11}. However, this approach introduces two additional sources of error to the variational approximation: It employs an approximate solution for the variational distribution of the auxiliary variables, and it provides an approximation for the posterior distribution of the extended parameter set, whose marginal does not necessarily minimize the Kullback-Leibler divergence with respect to the original posterior distribution. Instead, we relax the evidence lower bound following the approach of \cite{Jaakkola&Jordan00}. To this end, using a first order Taylor expansion around $\gamma_{\phi(\xi_i,\xi_j)}\in\Re,$ it is easy to show that 
\begin{multline*}
p(\mathcal{Y}\mid \Theta,\boldsymbol{\xi})\geq \tilde{p}(\mathcal{Y}\mid \Theta,\boldsymbol{\xi},\Gamma) \equiv \prod_{i=1}^{I-1}\prod_{j=i+1}^I\exp\left\{y_{i,j}\theta_{\phi(\xi_i,\xi_j)}\right\} S\left(\gamma_{\phi(\xi_i,\xi_j)}\right)\\
\times\exp\left\{\frac{-\left(\theta_{\phi(\xi_i,\xi_j)}+\gamma_{\phi(\xi_i,\xi_j)}\right)}{2}+\lambda\left(\gamma_{\phi(\xi_i,\xi_j)}\right)\left(\theta_{\phi(\xi_i,\xi_j)}^2-\gamma_{\phi(\xi_i,\xi_j)}^2\right)\right\},
\end{multline*}
where $\lambda(x)=\frac{1}{2x}\left(S(x)-\frac{1}{2}\right),$ $S(x)=(1+\exp\{-x\})^{-1},$ and the equality holds whenever $\theta^2_{k,l}=\gamma^2_{k,l}$ for all $k\leq l.$ Then, the relaxed lower bound is given by 
\begin{multline*}
\tilde{F}(q,\mathcal{Y})\equiv \textsf{E}_{q(\Theta,\boldsymbol{\xi})}\left[\log \tilde{p}(\mathcal{Y}\mid \Theta, \boldsymbol{\xi}, \Gamma) \right]+\textsf{E}_{q(\Upsilon)}\left[\log p(\Upsilon) \right]+H[q] \\ \leq \textsf{E}_{q(\Theta,\boldsymbol{\xi})}\left[\log p(\mathcal{Y}\mid \Theta,\boldsymbol{\xi}) \right]+\textsf{E}_{q(\Upsilon)}\left[\log p(\Upsilon) \right]+H[q]=F(q,\mathcal{Y}).
\end{multline*}
Maximization of $\tilde{F}(q,\mathcal{Y})$ leads to an approximation of the variational distribution for the community parameters with the form 
\begin{multline}\label{vartheta}
\log \tilde q(\theta_{k,l})= -\frac{1}{2}\left\{2\lambda(\gamma_{k,l})\textsf{E}_{q(\boldsymbol{\xi})}\left[n_{k,l}\right]+\textsf{E}_{q(\sigma^2)}\left[\frac{1}{\sigma^2}\right]\right\}\theta_{k,l}^2 \\
+\left\{\textsf{E}_{q(\boldsymbol{\xi})}\left[s_{k,l}\right]-\frac{\textsf{E}_{q(\boldsymbol{\xi})}\left[n_{k,l}\right]}{2}+\textsf{E}_{q(\sigma^2)}\left[\frac{1}{\sigma^2}\right]\textsf{E}_{q(H,\boldsymbol{\zeta})}\left[\eta_{\phi(\zeta_k,\zeta_l)}\right]\right\}\theta_{k,l}+C,
\end{multline}
a Gaussian kernel, while maximizing $\textsf{E}_{q(\Theta,\boldsymbol{\xi})}\left[\log \tilde{p}(\mathcal{Y}\mid \Theta,\boldsymbol{\xi},\Gamma) \right]$ gives that the optimal auxiliary parameters satisfy $\gamma_{k,l}^2=\textsf{E}_{q(\theta_{k,l})}\left[\theta_{k,l}^2\right].$

The  variational distributions for most of the rest of the parameters in the model take the same form as their full conditional counterpart. The expectations involved in the calculation of the variational parameters can then be found in closed form and a (iterative) variational Bayes algorithm can be implemented. The only notable exception is given by the variational distribution of the concentration parameters $\alpha$ and $\beta$, since the Gamma distribution is not conjugate in this case. However, using a first order approximation to $\log\Gamma(x)$ we are able to approximate the variational distributions $q^\star(\alpha)$ and $q^\star(\beta)$ with Gamma distributions closely related to those of \cite{Escobar&West95}. We leave the details to Appendix \ref{Bvarmulti}.

\section{Illustrations}
\label{ilust}

\begin{figure}[!htb]
\centering
\includegraphics[width=.6\textwidth]{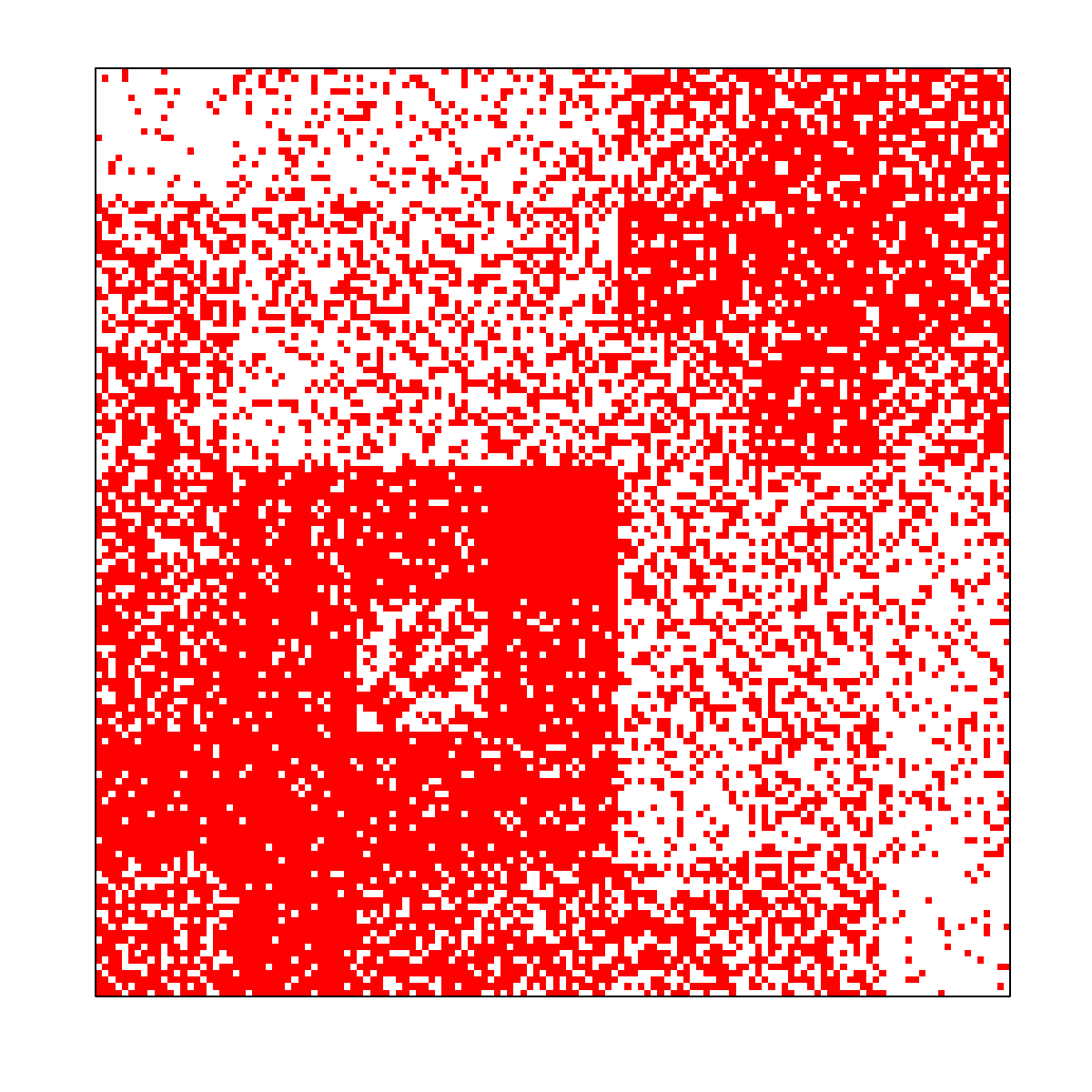} 
\caption{Image representation of the adjacency matrix. Here, actors in the network are placed along the horizontal and vertical axis. $y_{i,j}=1$ is represented by a red dot, while a lack of interaction is shown in white.}
\label{YD7}
\end{figure}

In this section, we illustrate the model and compare the performance of the two algorithms described in Section \ref{psml}. As a first step, we make use of the simulated dataset shown in Figure \ref{YD7}, this network is constructed with $I=140$ individuals evenly split into $K^\star=7$ communities. In turn, the network is split into $R^\star=2$ supercommunities formed by the first four and the last three communities respectively.

\begin{figure}[!htb]
\centering
\includegraphics[width=\textwidth]{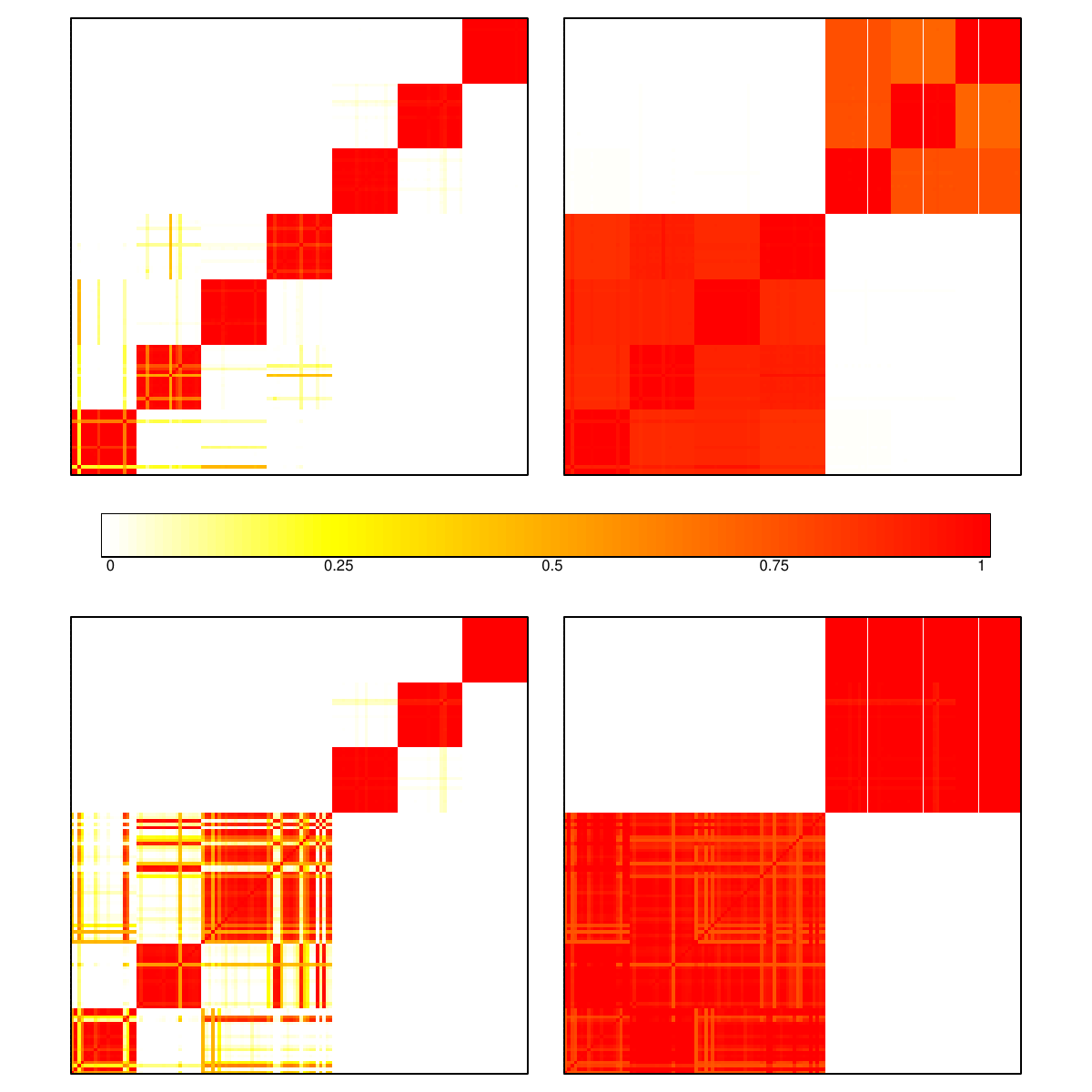} 
\caption{Community estimates for simulated data. \textbf{Top:} Monte Carlo estimates of pairwise posterior probabilities of same community, $\textsf{Pr}(\xi_i=\xi_j),$ (left), and supercomunity,  $\textsf{Pr}(\zeta_{\xi_i}=\zeta_{\xi_j}),$ (right). \textbf{Bottom} Variational approximations  $q(\xi_i=\xi_j),$ (left), and $q(\zeta_{\xi_i}=\zeta_{\xi_j})$ (right).}
\label{rD7}
\end{figure}

Choosing $K=20$ and $R=2,$ the top two panels of Figure \ref{rD7} show the MCMC posterior pairwise co-membership probabilities for communities and supercommunities respectively. That is, for every pair $(i,j)$ the left figure shows $\textsf{Pr}(\xi_i=\xi_j)$ while the right panel shows $\textsf{Pr}(\zeta_{\xi_i}=\zeta_{\xi_j}).$ From here, it can be observed that the model is capable of recovering the underlying community structure with little uncertainty in both levels. In turn, the bottom panels of Figure \ref{rD7} show the respective results obtained using the variational approximation.  In this case the algorithm is able to learn most of the structure on the data, although it is worthwhile noticing the higher levels of uncertainty and the fact that, in this particular solution, communities three and four are not well discerned.

\begin{figure}[!htb]
\centering
\includegraphics[width=.6\textwidth]{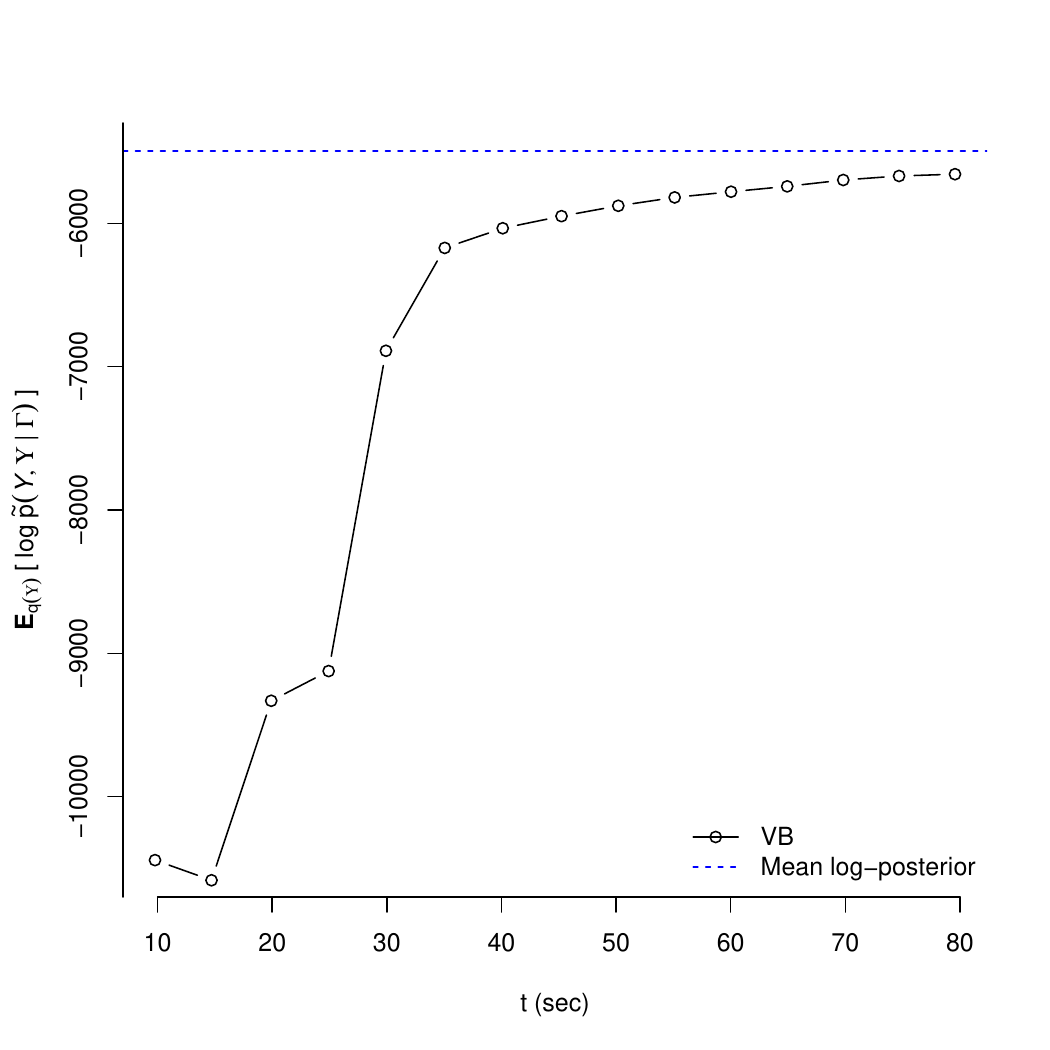} 
\caption{Evolution of the lower bound as a function of execution time.}
\label{boundD7}
\end{figure}

Now, notice that the multimodality of the problem makes both the MCMC and the variational approximation algorithms susceptible to the choice of the initial conditions. For this reason, the results shown in Figure \ref{rD7} correspond to those obtained from the run with the highest mean posterior likelihood and evidence lower bound out of 32 runs. This does not affect the computational efficiency of either algorithm as multiple runs are done in parallel. Figure \ref{boundD7} shows the evolution of the evidence lower bound as a function of execution time, and the mean posterior likelihood from the MCMC after $100,000$ posterior samples have been obtained in approximately $4,500$ seconds (75 minutes).

As a second illustration, we consider the co-authorship network from \cite{Newman06}. In this network, the vertices represent $I=379$ authors of scientific papers in the field of network science, and an edge between two vertices indicates that the authors have collaborated on at least one of the 914 publications included in the network. This network is a subset of a larger network constructed from the bibliographies of the two reviews \cite{Newman03} and \cite{Boccalettietal06}.

%\begin{figure}[!htb]
%\centering
%\includegraphics[width=.6\textwidth]{YNS.pdf} 
%\caption{Adjacency matrix for the collaboration network of \cite{Newman06}. This network is a subset of a larger network constructed from the bibliographies of the two reviews \cite{Newman03} and \cite{Boccalettietal06}. }
%\label{YNS}
%\end{figure} 

\begin{figure}[!htb]
\centering
\includegraphics[width=\textwidth]{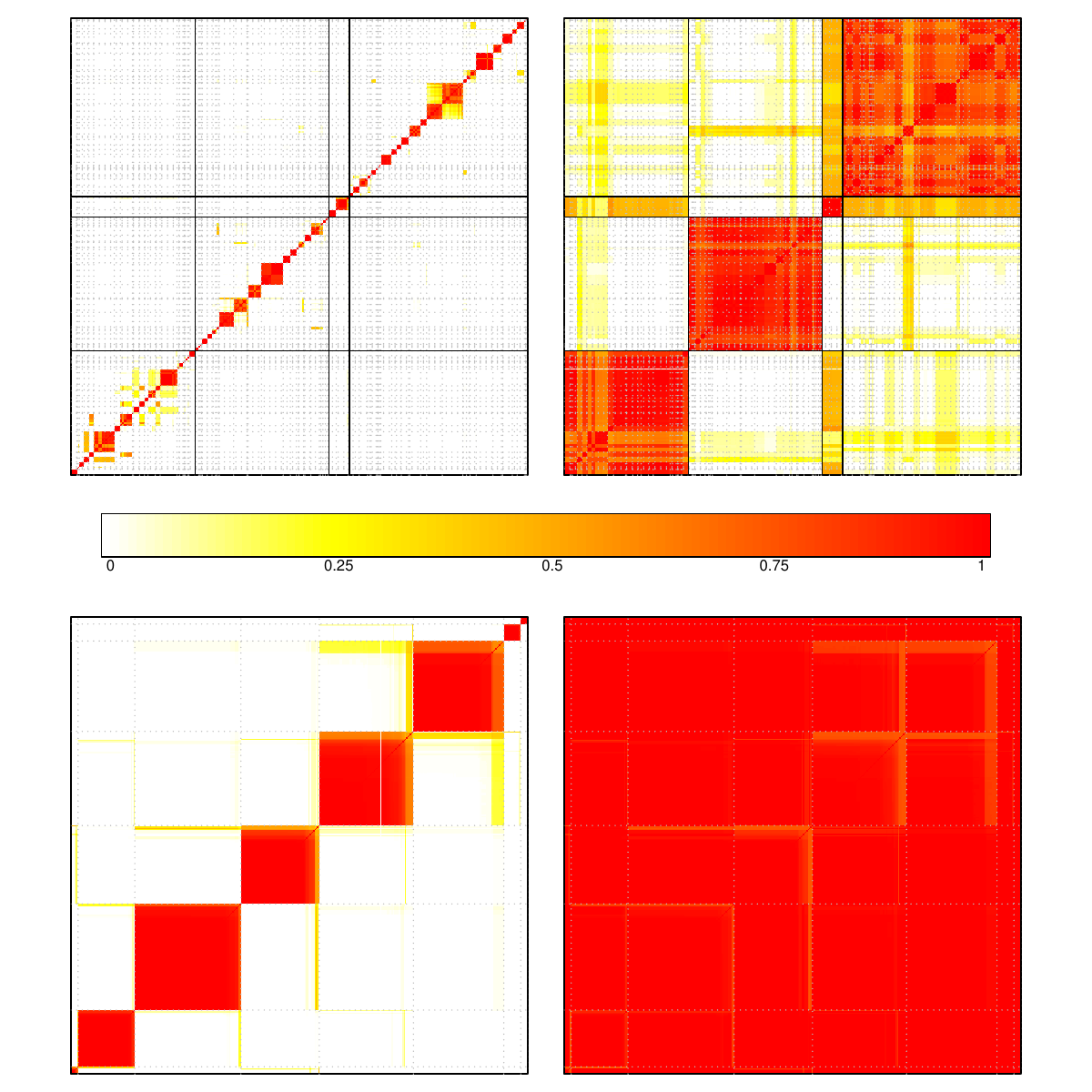} 
\caption{Community estimates for collaboration network. \textbf{Top:} Monte Carlo estimates of pairwise posterior probabilities of same community, $\textsf{Pr}(\xi_i=\xi_j),$ (left), and supercomunity,  $\textsf{Pr}(\zeta_{\xi_i}=\zeta_{\xi_j}),$ (right). \textbf{Bottom} Variational approximations  $q(\xi_i=\xi_j),$ (left), and $q(\zeta_{\xi_i}=\zeta_{\xi_j})$ (right).}
\label{rNS}
\end{figure}

Figure \ref{rNS} shows the structure recovered with $K=100$ and $R=15$ from the MCMC (top) and, with a different ordering of the vertices, for the variational Bayes (bottom) for communities (left) and supercommunities (right) respectively. From these plots it can be seen that, in this case, the inferred structure is significantly different between the two methods. In particular, with the MCMC a much larger number of smaller communities are found, with most of the clustering occurring in the second level, while the variational approximation finds four larger communities but does not capture any hierarchical structure.

\begin{figure}[!htb]
\centering
\includegraphics[width=.6\textwidth]{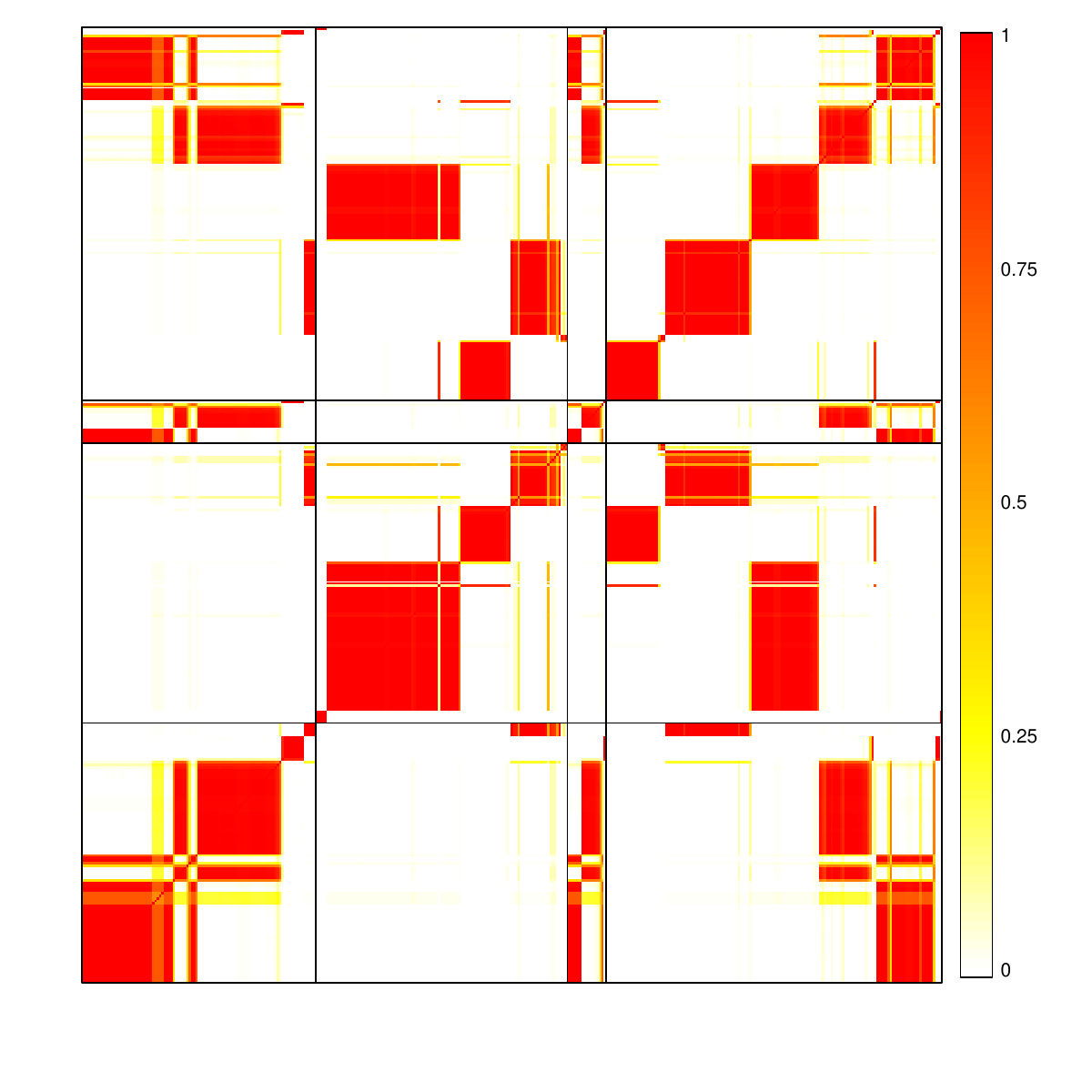} 
\caption{Overlap in community structure between obtained under the MCMC and variational algorithms. Colors correspond to the variational probabilities of  same community, while the ordering is taken to represent the hierarchical community structure from the MCMC.}
\label{rCrossNS}
\end{figure}

In order to asses the proximity of these two solutions, Figure \ref{rCrossNS} presents the overlap between the supercommunities obtained from the MCMC and the communities found by the variational algorithm.  That is, the variational estimates for the the first level pairwise co-clustering probabilities are shown displayed under the optimal ordering obtained with the MCMC. From this figure, it is interesting to note that although the communities from the variational algorithm are broken into different supercommunities in the MCMC, a good proportion of the vertices remain together.

\begin{figure}[!htb]
\centering
\includegraphics[width=\textwidth]{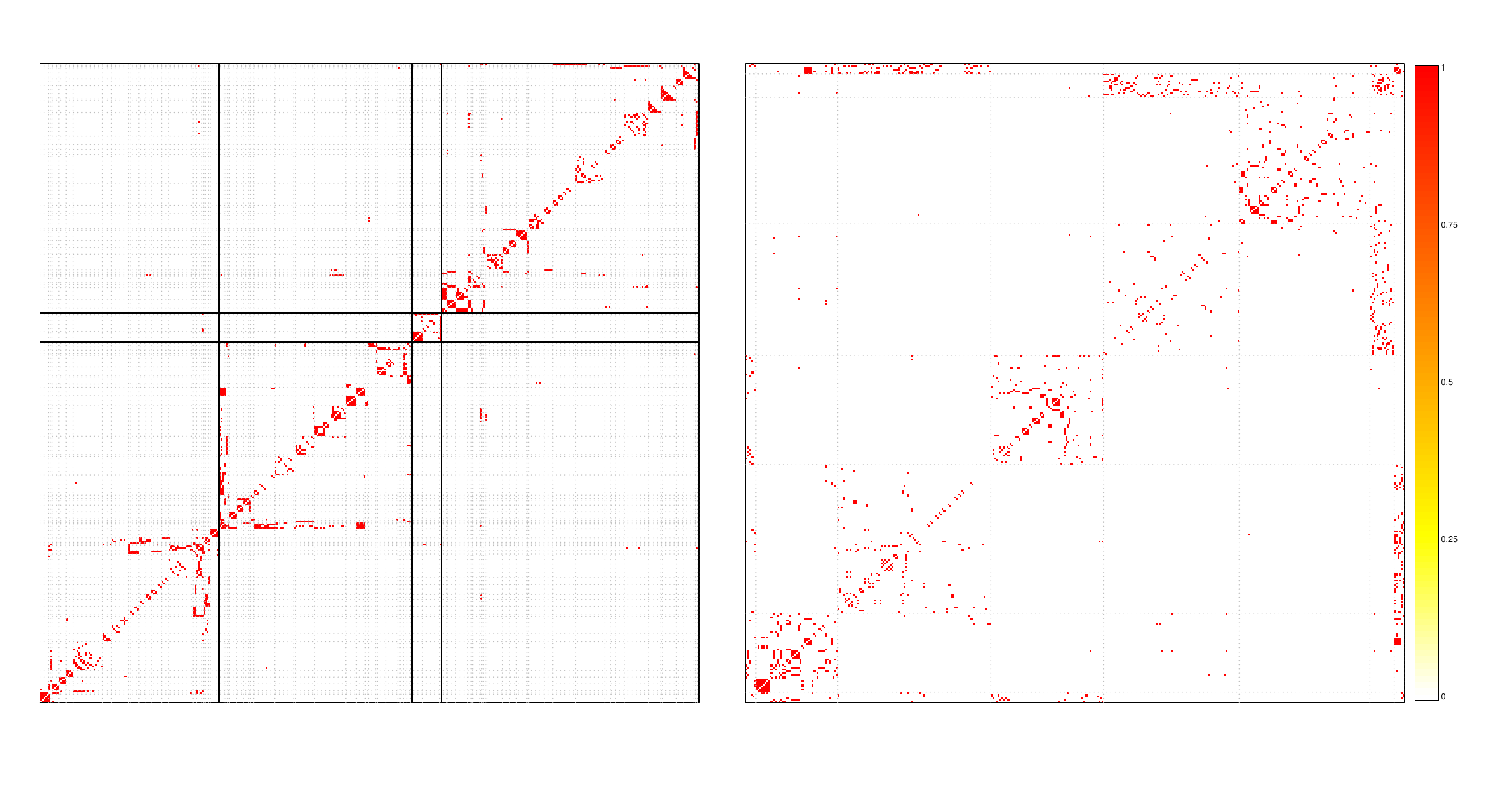} 
\caption{ Adjacency matrix of the collaboration network ordered with respect to MCMC (left) and variational (right) community structure.}
\label{YreordNS}
\end{figure}

\begin{figure}[!htb]
\centering
\includegraphics[width=.6\textwidth]{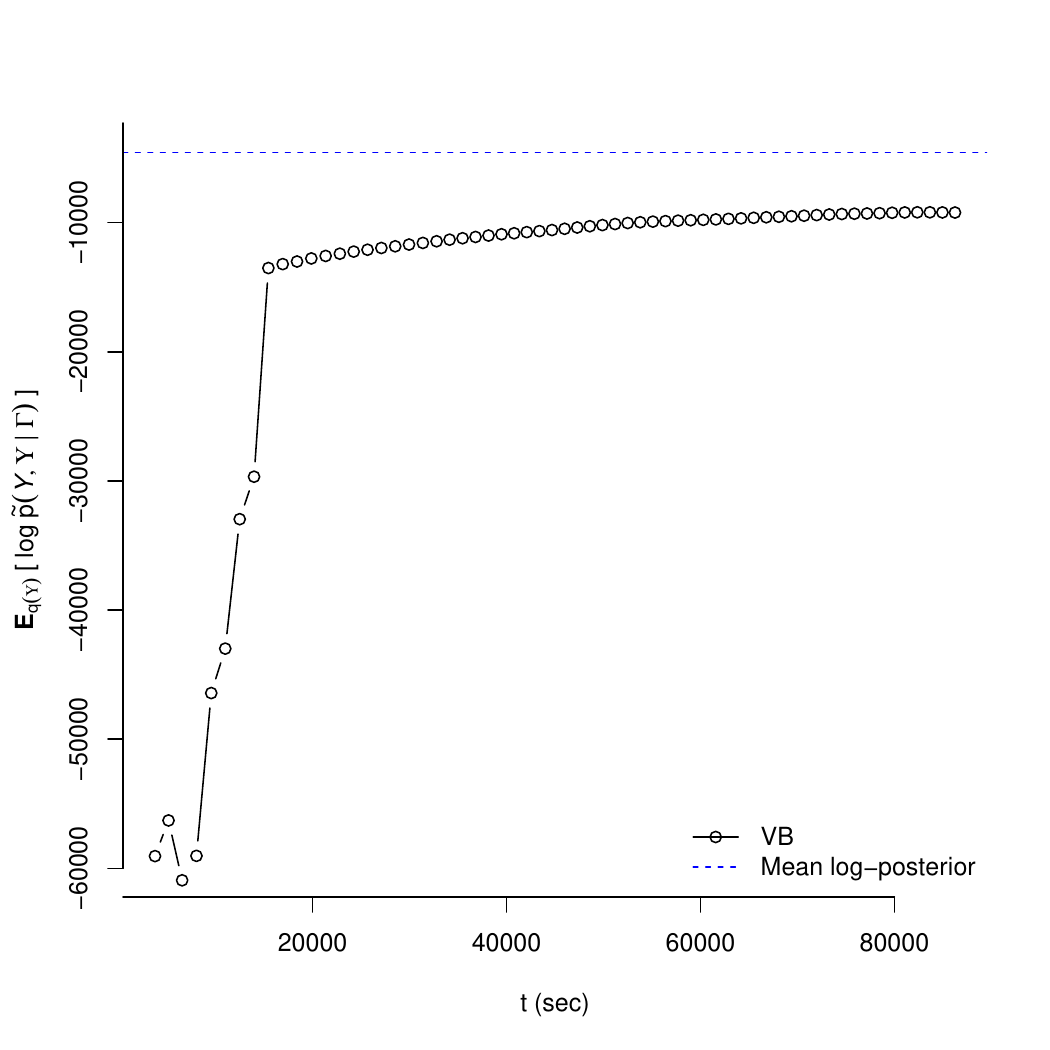} 
\caption{Evolution of the lower bound as a function of execution time. }
\label{boundNS}
\end{figure}

In turn, Figure \ref{YreordNS} shows the the adjacency matrix permuted to show the corresponding community structure under the MCMC (left) and variational Bayes (right). From this figure, it can be seen all four supercommunities recovered by the model trough the MCMC are highly assortative, while the multiple communities found within each supercommunity exhibit a slightly higher propensity of interaction. Instead, the larger communities found by the variational algorithm display a mixture of assortative and disassortative groups in the network.
Finally, Figure \ref{boundNS} shows the evolution of the evolution of the ELBO with respect to execution time, along the mean posterior likelihood of the $100,000$ posterior realizations from the MCMC obtained in approximately $153,000$ seconds $(42 hours).$

\section{Discussion}
\label{disc}

In this paper, we introduce a hierarchical extension of the stochastic blockmodel to identify multilevel community structures in networks. We also present a Markov chain Monte Carlo (MCMC) and a variational Bayes algorithm to fit the model and obtain approximate posterior inference. Through simulated and real datasets, we demonstrate that the model successfully identifies communities and supercommunities when they exist in the data. Additionally, we observe that the model returns a single supercommunity when there is no evidence of multilevel community structure.
As expected in the case of the single-level stochastic blockmodel, we observe that the MCMC algorithm consistently outperforms its variational Bayes counterpart. Therefore, we recommend using MCMC whenever the network size allows for computational feasibility.

Although our results in terms of inference for the community structure have shown
to be robust to the choice of prior for the scale parameters in the model, others have argued against the use of the Inverse gamma as a non-informative distribution \citep{Gelman2006}. For this reason, a possible direction for future research is testing an alternative prior distribution for these parameters, such as the scale Beta2 of \cite{Perez2017}.

Finally, note that for simplicity, this paper focuses on undirected and binary networks. However, generalizations to undirected and/or count networks are straightforward, requiring the removal of symmetry constraints in the community parameters and a change in the form of the kernel, respectively.

\section*{Statements and Declarations}

The authors declare that they have no known competing financial interests or personal relationships that could have appeared to influence the work reported in this article.

%\nocite{*}
\bibliography{references.bib}
\bibliographystyle{apalike}

\appendix

\section{Details for derivation of expected number of clusters in the multilevel stochastic blockmodel}\label{AENC}

Direct calculation from the distribution of $K^\star$ yields that 
\begin{equation*}
\textsf{E}\left[K^\star\right]=\alpha\left\{\Psi(\alpha+I)-\Psi(\alpha)\right\}\approx \alpha\log\left(\frac{\alpha+I}{\alpha}\right),
\end{equation*}
and
\begin{equation*}
\textsf{Var}\left[K^\star\right]=\alpha\left\{\Psi(\alpha+I)-\Psi(\alpha)\right\}+\alpha^2\left\{\Psi^\prime(\alpha+I)-\Psi^\prime(\alpha)\right\}\approx \alpha\log\left(\frac{\alpha+I}{\alpha}\right).
\end{equation*}

Therefore, using the law of iterated expectations, 
\begin{equation*}
\textsf{E}\left[R^\star\right]=\textsf{E}\left[\textsf{E}\left[R^\star\mid K^\star\right]\right]\approx\textsf{E}\left[\beta\log\left(\frac{\beta+K^\star}{\beta}\right)\right]=\beta\textsf{E}\left[\log\left(\frac{\beta+K^\star}{\beta}\right)\right].
\end{equation*}

And applying a second order Taylor approximation yields the desired result: 
\begin{align*} 
\textsf{E}\left[R^\star\right]\approx &\beta \left\{\log\left(\textsf{E}\left[\frac{\beta+K^\star}{\beta}\right]\right)-\frac{1}{2}\frac{\textsf{Var}\left[\frac{\beta+K^\star}{\beta}\right]}{\textsf{E}^2\left[\frac{\beta+K^\star}{\beta}\right]}\right\}\\
& = \beta \left\{\log\left(\frac{\beta+\textsf{E}\left[K^\star\right]}{\beta}\right)-\frac{1}{2}\frac{\frac{1}{\beta^2}\textsf{Var}\left[K^\star\right]}{\left(\frac{\beta+\textsf{E}\left[K^\star\right]}{\beta}\right)^2}\right\}\\
& \approx \beta \left\{\log\left(\frac{\beta+\alpha\log\left(\frac{\alpha+I}{\alpha}\right)}{\beta}\right)-\frac{1}{2}\frac{\alpha\log\left(\frac{\alpha+I}{\alpha}\right)}{\left(\beta+\alpha\log\left(\frac{\alpha+I}{\alpha}\right)\right)^2}\right\}.
\end{align*}

\section{Details for MCMC algorithm for the multilevel stochastic blockmodel} \label{AMCMCmult}

The full conditionals distributions can be derived from \eqref{postml}.  As a first step, let $\Upsilon_{-\theta_{k,l}}$ be the set of all parameters in the model with the exception of $\theta_{k,l}.$ Then, for $1\leq k\leq l\leq K,$ 
\begin{equation*}
p(\theta_{k,l}\mid\Upsilon_{-\theta_{k,l}},\Gamma,\mathcal{Y}) 
\propto  \exp\left\{-\frac{1}{2}\left(\gamma_{k,l}+\frac{1}{\sigma^2}\right)\theta_{k,l}^2+\left(s_{k,l}-\frac{n_{k,l}}{2}+\frac{\eta_{\phi\left(\zeta_k,\zeta_l\right)}}{\sigma^2}\right)\theta_{k,l}\right\},
\end{equation*}
which can be identified as a Gaussian kernel $\textsf{N}(\mu^\star_{\theta_{k,l}},\sigma^{2^\star}_{\theta_{k,l}})$ with mean parameter given by $\mu^\star_{\theta_{k,l}}=\left(\gamma_{k,l}+\frac{1}{\sigma^2}\right)^{-1}\left(s_{k,l}-\frac{n_{k,l}}{2}+\frac{\eta_{\phi\left(\zeta_k,\zeta_l\right)}}{\sigma^2}\right)$ and variance $\sigma^{2^\star}_{\theta_{k,l}}=\left(\gamma_{k,l}+\frac{1}{\sigma^2}\right)^{-1}.$

Now, for the auxiliary variables,
\begin{equation*}
p(\gamma_{k,l}\mid \Upsilon, \Gamma_{-(k,l)},\mathcal{Y}) \propto \exp\left\{-\frac{\theta_{k,l}^2}{2}\gamma_{k,l}\right\}\pi(\gamma_{k,l}),
\end{equation*}
that remains in the Polya-Gamma family, specifically, $\gamma_{k,l}\mid \Upsilon, \Gamma_{-(k,l)},\mathcal{Y}\sim\textsf{PG}(n_{k,l},\theta_{k,l}),$  and, thus, can be sampled using the approach proposed by \cite{Polsonetal13} that builds on \cite{Devroye09}.

In the case of the community indicators, 
\begin{equation*}\label{fcxi}
Pr(\xi_{i}=k\mid \Upsilon_{-\xi_i},\Gamma,\mathcal{Y})\propto w_{k}\prod_{\substack{j=1 \\ j\neq i}}^I\frac{(\exp\{\theta_{\phi\left(\xi_j,k\right)}\})^{y_{\phi\left(\xi_j,k\right)}}}{1+\exp\{\theta_{\phi\left(\xi_j,k\right)}\}},
\end{equation*}
which is a Categorical distribution.

For the first level variance parameter $\sigma^2,$ 
\begin{equation*}
p(\sigma^2\mid,\Upsilon_{-\sigma^2},\Gamma,\mathcal{Y}) \propto (\sigma^2)^{-\left(\frac{1}{4}K(K+1)+\alpha_\sigma+1\right)}\exp\left\{\frac{-\left[\frac{1}{2}\sum_{k=1}^K\sum_{l=k}^K(\theta_{k,l}-\eta_{\phi\left(\zeta_k,\zeta_l\right)})^2+\beta_\sigma\right]}{\sigma^2}\right\},
\end{equation*}
which is an $\textsf{IG}(\alpha_\sigma^{\star},\beta_\sigma^{\star})$ with $\alpha_\sigma^{\star}=\alpha_\sigma+\frac{1}{2}K(K+1)$ and $\beta_\sigma^{\star}=\frac{1}{2}\sum_{k=1}^K\sum_{l=k}^K(\theta_{k,l}-\eta_{\phi\left(\zeta_k,\zeta_l\right)})^2+\beta_\sigma.$ Notice, however, that it may be the case that some communities have no subjects assigned to them. Therefore, mixing in this sampling scheme can be improved by defining  
\begin{equation*}
x_{k,l}=\begin{cases} 1,\quad \text{if } n_{k,l}\geq1;\\
0, \quad\text{otherwise,} \end{cases}
\end{equation*}
for $1\leq k\leq l \leq K$ and $N=\sum_{k=1}^K\sum_{l=k}^K x_{k,l},$ and sample $\sigma^2$ from an $\textsf{IG}(\alpha^{\star\star},\beta^{\star\star})$ with parameters $\alpha_\sigma^{\star\star}=\frac{N+2\alpha_\sigma}{2}$ and $\beta_\sigma^{\star\star}=\frac{1}{2}\sum_{k=1}^K\sum_{l=k}^K(\theta_{k,l}-\eta_{\phi\left(\zeta_k,\zeta_l\right)})^2x_{k,l}+\beta_\sigma.$

For the elements of $H,$ 
\begin{equation*}
p(\eta_{r,s}\mid\Upsilon_{-\eta_{r,s}},\Gamma,\mathcal{Y}) 
\propto \exp\left\{-\frac{1}{2}\left(\frac{m_{r,s}}{\sigma^2}+\frac{1}{\tau^2}\right)\eta_{r,s}^2+\left(\frac{t_{r,s}}{\sigma^2}+\frac{\mu}{\tau^2}\right)\eta_{r,s}\right\},
\end{equation*}
where $m_{r,s}$ and $t_{r,s}$ play the role of $n_{k,l}$ and $s_{k,l}$ respectively in the second level of the hierarchy. That is, $t_{r,s}=\sum_{\mathcal{T}_{r,s}}\theta_{k,l}$ while $m_{r,s}=\sum_{\mathcal{T}_{r,s}}1,$ with the sum taken over the set $\mathcal{T}_{r,s}=\left\{(k,l):\; k\leq l,\; (r,s)=\phi\left(\zeta_k,\zeta_l\right)\right\}.$ Thus, $\eta_{r,s}$  can be sampled from of a $\textsf{N}(\mu^\star_{\eta_{r,s}},\tau^{2^\star}_{\eta_{r,s}})$ with mean $\mu^\star_{\eta_{r,s}}=\left(\frac{m_{r,s}}{\sigma^2}+\frac{1}{\tau^2}\right)^{-1}\left(\frac{t_{r,s}}{\sigma^2}+\frac{\mu}{\tau^2}\right)$ and variance $\tau^{2^\star}_{\eta_{r,s}}=\left(\frac{m_{r,s}}{\sigma^2}+\frac{1}{\tau^2}\right)^{-1}.$

Similar to the case of $\boldsymbol{\xi},$ the full conditional for $\boldsymbol{\zeta}$ satisfies that 
\begin{equation*}
p(\boldsymbol{\zeta}\mid,\Upsilon_{-\boldsymbol{\zeta}},\Gamma,\mathcal{Y})\propto p(\boldsymbol{\zeta}\,|\,\boldsymbol{v})p(\Theta\,|\,H,\boldsymbol{\zeta},\sigma^2).
\end{equation*}
That is, for every $r=1,2,\ldots,R,$ and every $k=1,2,\ldots,K,$
\begin{equation*}\label{fczetaml}
Pr(\zeta_{k}=r\mid\Upsilon_{-\zeta_k},\Gamma,\mathcal{Y})\propto v_{r} \exp\left\{-\frac{1}{2\sigma^2}\sum_{l=1}^K(\theta_{\phi\left(\xi_j,k\right)}-\eta_{\phi\left(r,\zeta_l\right)})^2\right\}.
\end{equation*}

The full conditional for $\mu$ is given by  
\begin{equation*}
p(\mu\mid,\Upsilon_{-\mu},\Gamma,\mathcal{Y}) 
\propto \exp\left\{-\frac{1}{2}\left(\frac{\frac{1}{2}R(R+1)}{\tau^2}+\frac{1}{\sigma^2_\mu}\right)\mu^2+\mu\left(\frac{\sum_{r=1}^R\sum_{s=r}^R\eta_{r,s}}{\tau^2}+\frac{\mu_\mu}{\sigma^2_\mu}\right)\right\},
\end{equation*}
which is then sampled from a Gaussian $\textsf{N}(\mu^{\star\star}_{\mu},\sigma^{2^{\star\star}}_{\mu})$ with variance $\sigma^{2^{\star\star}}_{\mu}=\left(\frac{M}{\tau^2}+\frac{1}{\sigma^2_\mu}\right)^{-1},$ and mean $\mu^{\star\star}_{\mu}=\left(\frac{M}{\tau^2}+\frac{1}{\sigma^2_\mu}\right)^{-1}\left(\frac{\sum_{r=1}^R\sum_{s=r}^R\eta_{r,s}z_{r,s}}{\tau^2}+\frac{\mu_\mu}{\sigma^2_\mu}\right),$ and where, analogously to the first level, $M=\sum_{r=1}^R\sum_{s=r}^Rz_{r,s}$ and, for $1\leq r\leq s \leq R$,
\begin{equation*}
z_{r,s}=\begin{cases} 1,\quad \text{if } m_{r,s}\geq1;\\
0, \quad\text{otherwise.} \end{cases}
\end{equation*}

As in the case of $\sigma^2,$ for $\tau^2$,  
\begin{equation*}
p(\tau^2\mid,\Upsilon_{-\tau^2},\Gamma,\mathcal{Y}) \propto (\tau^2)^{-\left(\frac{1}{4}R(R+1)+\alpha_\tau+1\right)}\exp\left\{\frac{-\left[\frac{1}{2}\sum_{r=1}^R\sum_{s=r}^R(\eta_{r,s}-\mu)^2+\beta_\tau\right]}{\tau^2}\right\},
\end{equation*}
which is sampled from $\textsf{IG}(\alpha_\tau^{\star\star},\beta_\tau^{\star\star})$ with $\alpha_\tau^{\star\star}=\frac{M+2\alpha_\tau}{2}$ and $\beta_\tau^{\star\star}=\frac{1}{2}\sum_{r=1}^R\sum_{s=r}^R(\eta_{r,s}-\mu)^2z_{r,s}+\beta_\tau.$

Now, the weights can be sampled from 
\begin{equation*}
p(\boldsymbol{w}\mid \Upsilon_{-\boldsymbol{w}},\Gamma,\mathcal{Y}) \propto \prod_{k=1}^K w_k^{\frac{\alpha}{K}+n_k-1},
\end{equation*}
a Dirichlet with parameter vector $\left(\frac{\alpha}{K}+n_1,\frac{\alpha}{K}+n_2,\ldots,\frac{\alpha}{K}+n_K\right)$ and, similarly, 
\begin{equation*}
\boldsymbol{v}\mid \Upsilon_{-\boldsymbol{v}},\Gamma,\mathcal{Y}\sim \textsf{Dir}\left(\frac{\beta}{R}+m_1,\frac{\beta}{R}+m_2,\ldots,\frac{\beta}{R}+m_R\right).
\end{equation*}

Finally, for the concentration parameters, 
\begin{equation*}
p(\alpha\mid \Upsilon_{-\alpha},\Gamma,\mathcal{Y}) \propto \frac{\Gamma(\alpha)}{\left[\Gamma\left(\frac{\alpha}{K}\right)\right]^K} \alpha^{\alpha_\alpha-1} \exp\{-\beta_\alpha \alpha\},
\end{equation*}
that does not lead to a closed form. Thus, $\alpha$ could be sampled using a Metropolis-Hastings step with, for example, a random walk on the log scale. Alternatively, by recognizing that for large enough $K$ this distribution approximates the full conditional for the concentration parameters of a Dirichlet process, $\alpha$ can be sampled from a mixture of Gammas borrowing from \cite{Escobar&West95}. The case of $\beta$ is analogous for large enough $R.$

\section{Details for the variational Bayes algorithm for the multilevel stochastic blockmodel}\label{Bvarmulti}

The optimal variational distribution for the community indicators is Categorical satisfying:
\begin{multline*}
\log q^\star(\xi_i=k)=\textsf{E}_{q(w_k)}\left[\log w_k\right]+\sum_{j\neq i}y_{\phi(i,j)}\sum_{l=1}^K\textsf{E}_{q(\theta_{\phi(k,l)})}\left[\theta_{\phi(k,l)}\right]q^\star(\xi_j=l)\\
+\sum_{j\neq i}\sum_{l=1}^K\textsf{E}_{q(\theta_{\phi(k,l)})}\left[\log\left(1+\exp\left\{\theta_{\phi(k,l)}\right\}\right)\right]q^\star(\xi_j=l)+C.
\end{multline*}

For the first level variance, 
\begin{multline*}
\log q^\star(\sigma^2)= -\left(\frac{1}{4}K(K+1)+\alpha_\sigma+1\right)\log \sigma^2\\-\frac{\frac{1}{2}\sum_{k=1}^K\sum_{l=k}^K\textsf{E}_{q(\Theta,H,\zeta)}\left[(\theta_{k,l}-\eta_{\phi(\zeta_k,\zeta_l)})^2\right]+\beta_\sigma}{\sigma^2}+C,
\end{multline*}
which is readily identified as an Inverse Gamma distribution.

The distribution for the location parameters $H$ is also approximated by a Gaussian, in this case  
\begin{multline*}
\log q^\star(\eta_{r,s})= -\frac{1}{2}\left\{\textsf{E}_{q(\sigma)}\left[\frac{1}{\sigma^2}\right]\textsf{E}_{q(\boldsymbol{\zeta})}\left[m_{r,s}\right]+\textsf{E}_{q(\tau)}\left[\frac{1}{\tau^2}\right]\right\}\eta_{r,s}^2\\
+\left\{\textsf{E}_{q(\sigma)}\left[\frac{1}{\sigma^2}\right]\textsf{E}_{q(\boldsymbol{\zeta})}\left[t_{r,s}\right]+\textsf{E}_{q(\tau)}\left[\frac{1}{\tau^2}\right]\textsf{E}_{q(\mu)}\left[\mu\right]\right\}\eta_{r,s}+C.
\end{multline*}

Similar to the case of $\boldsymbol{\xi},$ the supercommunity indicators follow a Categorical distribution with probabilities given by
\begin{equation*}
\log q^\star(\zeta_k=r)=\textsf{E}_{q(v_r)}\left[\log v_r\right]-\frac{1}{2}\textsf{E}_{q(\sigma)}\left[\frac{1}{\sigma^2}\right]\sum_{l\neq k}\textsf{E}_{q\left(\Theta,H,\boldsymbol{\zeta}\right)}\left[\left(\theta_{\phi(k,l)}-\eta_{\phi(\zeta_k,\zeta_l)}\right)^2\right]+C.
\end{equation*}

Now, for the second level mean parameter, 
\begin{multline*}
\log q^\star(\mu)= -\frac{1}{2}\left\{\frac{1}{2}R(R+1) \textsf{E}_{q(\tau)}\left[\frac{1}{\tau^2}\right]+\frac{1}{\sigma^2_\mu}\right\}\mu^2\\
+\left\{\textsf{E}_{q(\tau)}\left[\frac{1}{\tau^2}\right]\sum_{r=1}^R\sum_{s=r}^R\textsf{E}_{q(H)}\left[\eta_{r,s}\right]+\frac{\mu_\mu}{\sigma^2_\mu}\right\}\mu+C,
\end{multline*}
again, a Gaussian distribution, while the variational distribution for the variance parameter is an Inverse Gamma such that  
\begin{equation*}
\log q^\star(\tau^2)= -\left(\frac{1}{4}R(R+1)+\alpha_\tau+1\right)\log \tau^2-\frac{\frac{1}{2}\sum_{r=1}^R\sum_{s=r}^R\textsf{E}_{q(H,\mu)}\left[(\eta_{r,s}-\mu)^2\right]+\beta_\tau}{\tau^2}+C.
\end{equation*}

With respect to the weight vectors, it can be observed that 
\begin{equation*}
\log q^\star(\mathbf{w})= \sum_{k=1}^K\left(\textsf{E}_{q(\boldsymbol{\xi})}\left[n_{k}\right]+\frac{1}{K}\textsf{E}_{q(\alpha)}\left[\alpha\right]-1\right)\log w_k+C
\end{equation*}
and 
\begin{equation*}
\log q^\star(\mathbf{v})= \sum_{r=1}^R\left(\textsf{E}_{q(\boldsymbol{\zeta})}\left[m_{r}\right]+\frac{1}{R}\textsf{E}_{q(\beta)}\left[\beta\right]-1\right)\log v_r+C,
\end{equation*}
which remain in the Dirichlet distribution.

Finally, for the first level concentration parameter $\alpha,$
\begin{equation*}
\log q^\star(\alpha)=\log\Gamma(\alpha)-K\log\Gamma\left(\frac{\alpha}{K}\right)+\frac{1}{K}\sum_{k=1}^K\textsf{E}_{q(w_k)}\left[\log w_k\right]+(\alpha_\alpha-1)\log\alpha-\beta_\alpha\alpha+C,
\end{equation*}
which, using that $\log\Gamma(x)=(x-1)\log x+\mathcal{O}(\log x),$ can be approximated by 
\begin{equation*}
\log \tilde q(\alpha)=\log\alpha\left(\alpha_\alpha+K-1-1\right)-\alpha\left(\beta_\alpha-\frac{1}{K}\sum_{k=1}^K\left\{\textsf{E}_{q(w_k)}\left[\log w_k\right]-\log\frac{1}{K}\right\}\right)+C,
\end{equation*}
and, analogously, 
\begin{equation*}\label{varbeta}
\log \tilde q(\beta)=\log\beta\left(\alpha_\beta+R-1-1\right)-\beta\left(\beta_\beta-\frac{1}{R}\sum_{r=1}^R\left\{\textsf{E}_{q(v_r)}\left[\log v_r\right]-\log\frac{1}{R}\right\}\right)+C.
\end{equation*}

From equations \eqref{vartheta} to \eqref{varbeta} is possible to derive the complete algorithm with the respective optimal variational parameters as shown below
\begin{equation*}
\tilde q(\theta_{k,l})=\textsf{N}(m_{k,l},\nu_{k,l}),
\end{equation*}
with $m_{k,l}=\nu_{k,l}\left[\sum_{i=1}^{I-1}\sum_{j=i+1}^Iy_{i,j}\epsilon_{k,l}^{i,j}-\frac{1}{2}\sum_{i=1}^{I-1}\sum_{j=i+1}^I\epsilon_{k,l}^{i,j}+\frac{o}{p}\sum_{r=1}^R\sum_{s=r}^Rg_{r,s}\delta_{r,s}^{k,l}\right]$ and \\ $\nu_{k,l}=\left[\frac{1}{2}\sum_{i=1}^{I-1}\sum_{j=i+1}^I\epsilon_{k,l}^{i,j}\left(\frac{1-\exp\{-\gamma_{k,l}\}}{\gamma_{k,l}(1+\exp\{\gamma_{k,l}\})}\right)+\frac{o}{p}\right]^{-1},$ while $\gamma_{k,l}^2=\nu_{k,l}+m_{k,l}^2$
where 
\begin{equation*}
\epsilon_{k,l}^{i,j}=\begin{cases}q(\xi_i=k)q(\xi_j=l)+q(\xi_i=l)q(\xi_j=k),\quad &\text{ if } k\neq l;\\ q(\xi_i=k)q(\xi_j=l), \quad&\text{ if }k=l. \end{cases}
\end{equation*} 
and analogously, 
\begin{equation*}
\delta_{r,s}^{k,l}=\begin{cases}q(\zeta_k=r)q(\zeta_l=s)+q(\zeta_k=s)q(\zeta_l=r),\quad &\text{ if } r\neq s;\\ q(\zeta_k=l)q(\zeta_l=s), \quad&\text{ if }r=s. \end{cases}
\end{equation*}

While, for the community indicators 
\begin{equation*}
q^\star(\xi_i=k)=\varpi_{i,k},
\end{equation*}
with 
\begin{equation*}
\resizebox{\hsize}{!}{$\displaystyle \log \varpi_{i,k} = \boldsymbol{\Psi}\left(\psi_k\right)+\sum_{j\neq i}\left[y_{\phi(i,j)}\sum_{l=1}^Km_{\phi(k,l)}\varpi_{j,l}-\sum_{l=1}^K\left(\log\left(1+\exp\left\{m_{k,l}+\frac{1}{2}\nu_{k,l}\right\}\right)-\frac{\left(\exp\{\nu_{k,l}\}-1\right)\exp\{2m_{k,l}+\nu_{k,l}\}}{2\left(1+\exp\left\{m_{k,l}+\frac{1}{2}\nu_{k,l}\right\}\right)^2}\right)\varpi_{j,l}\right]+C$}
\end{equation*}
and where it has been made use of the fact that, from the second order Delta method, $\textsf{E}_{q(\theta_{k,l})\left[\log (1+\exp \theta_{k,l})\right]}\approx \log\left(1+\exp\left\{m_{k,l}+\frac{1}{2}\nu_{k,l}\right\}\right)-\frac{\left(\exp\{\nu_{k,l}\}-1\right)\exp\{2m_{k,l}+\nu_{k,l}\}}{2\left(1+\exp\left\{m_{k,l}+\frac{1}{2}\nu_{k,l}\right\}\right)^2}.$

In the case of the first level variance parameter, 
\begin{equation*}
q^\star(\sigma^2)=\textsf{IG}(o,p),
\end{equation*}
with \resizebox{.95\hsize}{!}{$\displaystyle p=\beta_\sigma+\frac{1}{2}\sum_{k=1}^K\sum_{l=k}^K\left(\nu_{k,l}+m_{k,l}^2\right)-\sum_{r=1}^R\sum_{s=r}^Rg_{r,s}\sum_{k=1}^K\sum_{l=k}^Km_{k,l}\delta_{r,s}^{k,l}+\frac{1}{2}\sum_{r=1}^R\sum_{s=r}^R\left(g_{r,s}^2+h_{r,s}^2\right)\sum_{k=1}^K\sum_{l=k}^K\delta_{r,s}^{k,l}$} and $o=\alpha_\sigma+\frac{1}{4}K(K+1).$

For the upper level, the location parameters satisfy
\begin{equation*}
q^\star(\eta_{r,s})=\textsf{N}(g_{r,s},h_{r,s}^2),
\end{equation*}
where $g_{r,s}=h_{r,s}^2\left[\frac{o}{p}\sum_{k=1}^K\sum_{l=k}^Km_{k,l}\delta_{r,s}^{k,l}+\frac{a}{b}c\right]$ and $h_{r,s}^2=\left[\frac{o}{p}\sum_{k=1}^K\sum_{l=k}^K\delta_{r,s}^{k,l}+\frac{a}{b}\right]^{-1},$ 
and the supercommunity indicators probabilities are such that
\begin{equation*}
q^\star(\zeta_k=r)=\varrho_{k,r},
\end{equation*}
with 
\begin{equation*}
\resizebox{\hsize}{!}{$\displaystyle \log \varrho_{k,r} = \boldsymbol{\Psi}\left(\varphi_r\right)-\frac{1}{2}\left(\frac{o}{p}\right)\sum_{l=1}^K\left[\nu_{\phi(k,l)}+m^2_{\phi(k,l)}-2m_{\phi(k,l)}\sum_{s=1}^Rg_{\phi(r,s)}\varrho_{l,s}+\sum_{s=1}^R\left(h_{\phi(r,s)}^2+g_{\phi(r,s)}^2\right)\right]+C.$}
\end{equation*}

In the case of the hyperparameters, the optimal variational distribution for the mean is  
\begin{equation*}
q^\star(\mu)=\textsf{N}(c,d^2),
\end{equation*}
where $c=d^2\left[\frac{a}{b}\sum_{r=1}^R\sum_{s=r}^Rg_{r,s}+\frac{\mu_\mu}{\sigma_\mu^2}\right]$ and $d^2=\left[\left(\frac{a}{b}\right)\frac{1}{2}R(R+1)+\frac{1}{\sigma_\mu^2}\right]^{-1},$ 
while for the variance 
\begin{equation*}
q^\star(\tau^2)= \textsf{IG}(a,b),
\end{equation*}
with $\displaystyle b=\beta_\tau+\frac{1}{2}\sum_{r=1}^R\sum_{s=r}^R\left(h_{r,s}^2+g_{r,s}^2\right)-c\sum_{r=1}^R\sum_{s=r}^Rg_{r,s}+\frac{1}{4}R(R+1)\left(d^2+c^2\right)$ 
\\ and $a=\alpha_\tau+\frac{1}{4}R(R+1).$

Lastly, for the weight parameters on the indicators side, 
\begin{equation*}
q^\star(\mathbf{w})=\textsf{Dir}(\boldsymbol{\psi}),
\end{equation*}
with $\displaystyle \psi_k=\frac{1}{K}\left(\frac{a_\alpha}{b_\alpha}\right)+\sum_{i=1}^I\varpi_{i,k},$ and 
\begin{equation*}
q^\star(\mathbf{v})=\textsf{Dir}(\boldsymbol{\varphi}),
\end{equation*}
with $\varphi_r=\frac{1}{R}\left(\frac{a_\beta}{b_\beta}\right)+\sum_{k=1}^K\varrho_{k,r},$
while their respective hyperparameters satisfy
\begin{equation*}
\tilde q(\alpha)=\textsf{G}(a_\alpha,b_\alpha),
\end{equation*}
with parameters $a_\alpha=\alpha_\alpha+K-1$ and $b_\alpha=\beta_\alpha-\frac{1}{K}\sum_{k=1}^K\left[\boldsymbol{\Psi}\left(\psi_k\right)-\boldsymbol{\Psi}\left(\sum_{l=1}^K\psi_l\right)-\log\left(\frac{1}{K}\right)\right],$
and
\begin{equation*}
\tilde q(\beta)=\textsf{G}(a_\beta,b_\beta),
\end{equation*}
where $a_\beta=\alpha_\beta+R-1$ and $\displaystyle b_\beta=\beta_\beta-\frac{1}{R}\sum_{r=1}^R\left[\boldsymbol{\Psi}\left(\varphi_r\right)-\boldsymbol{\Psi}\left(\sum_{s=1}^R\varphi_s\right)-\log\left(\frac{1}{R}\right)\right].$

Finally, it is found that the ELBO satisfies 
{\footnotesize
\begin{multline*}
\tilde{F}(q,\mathcal{Y})\approx \sum_{i<j}\left[y_{i,j}\sum_{k\leq l}m_{k,l}\epsilon_{k,l}^{i,j}-\sum_{k\leq l}\log \left(1+\exp\left\{-\gamma_{k,l}\right\}\right)\epsilon_{k,l}^{i,j}-\frac{1}{2}\left(m_{k,l}+\gamma_{k,l}\right)\epsilon_{k,l}^{i,j}\right.\\ \left. -\frac{1-\exp\left\{-\gamma_{k,l}\right\}}{4\gamma_{k,l}\left(1+\exp\left\{-\gamma_{k,l}\right\}\right)}\left(m_{k,l}^2+v_{k,l}-\gamma_{k,l}^2\right)\epsilon_{k,l}^{i,j}\right]
-\frac{K(K+1)}{4}\left(\log p-\boldsymbol{\Psi}(o)\right)\\-\frac{o}{2p}\sum_{k\leq l}\left[(m_{k,l}^2+v_{k,l})-2m_{k,l}\sum_{r\leq s}g_{r,s}\delta_{r,s}^{k,l}+\sum_{r\leq s}\left(g_{r,s}^2+h_{r,s}^2\right)\delta_{r,s}^{k,l}\right]
+\sum_{k=1}^K\left[\boldsymbol{\Psi}(\psi_k)\sum_{i=1}^I\varpi_{i,k}\right]\\-I\boldsymbol{\Psi}\left(\sum_{k=1}^K\psi_k\right)
+\alpha_\sigma\log \beta_\sigma -\log \Gamma(\alpha_\sigma)-(\alpha_\sigma-1)(\log p-\boldsymbol{\Psi}(o))-\beta_\sigma\frac{o}{p}
-\frac{R(R+1)}{4}\left(\log b-\boldsymbol{\Psi}(a)\right)\\ -\frac{a}{2b}\left[\sum_{r\leq s}\left(g_{r,s}^2+h_{r,s}^2\right)-2c\sum_{r\leq s}g_{r,s}+\frac{1}{2}R\left(R+1\right)\left(c^2+d^2\right)\right]
+\sum_{r=1}^R\left[\boldsymbol{\Psi}(\varphi_r)\sum_{k=1}^K\varrho_{k,r}\right]\\-K\boldsymbol{\Psi}\left(\sum_{r=1}^R\varphi_r\right)
-\frac{1}{2}\log(\sigma_\mu^2)-\frac{c^2+d^2-2c\mu_\mu+\mu_\mu^2}{\sigma_\mu^2}
+\alpha_\tau\log \beta_\tau -\log \Gamma(\alpha_\tau)-(\alpha_\tau-1)(\log b-\boldsymbol{\Psi}(a))\\-\beta_\tau\frac{a}{b}
+\log \Gamma\left(\frac{a_\beta}{b_\beta}\right)+\frac{1}{2}\boldsymbol{\Psi}_1\left(\frac{a_\beta}{b_\beta}\right)\frac{a_\beta}{b^2_\beta}-R\left[\log \Gamma\left(\frac{a_\beta}{Rb_\beta}\right)+\frac{1}{2}\boldsymbol{\Psi}_1\left(\frac{a_\beta}{Rb_\beta}\right)\frac{a_\beta}{R^2b^2_\beta}\right]\\ +\sum_{r=1}^R\left(\frac{a_\beta}{Rb_\beta}-1\right)\left(\boldsymbol{\Psi}(\varphi_r)-\boldsymbol{\Psi}\left(\sum_{s=1}^R\varphi_s\right)\right)
+\log \Gamma\left(\frac{a_\alpha}{b_\alpha}\right)+\frac{1}{2}\boldsymbol{\Psi}_1\left(\frac{a_\alpha}{b_\alpha}\right)\frac{a_\alpha}{b^2_\alpha}\\-K\left[\log \Gamma\left(\frac{a_\alpha}{Kb_\alpha}\right)+\frac{1}{2}\boldsymbol{\Psi}_1\left(\frac{a_\alpha}{Kb_\alpha}\right)\frac{a_\alpha}{K^2b^2_\alpha}\right]+\sum_{k=1}^K\left(\frac{a_\alpha}{Kb_\alpha}-1\right)\left(\boldsymbol{\Psi}(\psi_k)-\boldsymbol{\Psi}\left(\sum_{l=1}^K\psi_l\right)\right)\\
+\alpha_\alpha\log \beta_\alpha -\log \Gamma(\alpha_\alpha)+(\alpha_\alpha-1)(\boldsymbol{\Psi}(a_\alpha)-\log b_\alpha)-\beta_\alpha\frac{a_\alpha}{b_\alpha}
+\alpha_\beta\log \beta_\beta -\log \Gamma(\alpha_\beta)\\+(\alpha_\beta-1)(\boldsymbol{\Psi}(a_\beta)-\log b_\beta)-\beta_\beta\frac{a_\beta}{b_\beta}
+\frac{K(K+1)}{4}+\frac{1}{2}\sum_{k\leq l}\log(v_{k,l})
-\sum_{i=1}^I\sum_{k=1}^K\varpi_{i,k}\log \varpi_{i,k}\\
+o+\log p+\log \Gamma(o) -(1+o)\boldsymbol{\Psi}(o)
+\frac{R(R+1)}{4}+\frac{1}{2}\sum_{r\leq s}\log(h_{r,s}^2)
-\sum_{k=1}^K\sum_{r=1}^R\varrho_{k,r}\log \varrho_{k,r}\\
+\frac{1}{2}\left(1+\log d^2\right)
+a+\log b+ \log \Gamma(a)-(1+a)\boldsymbol{\Psi}(a)
+\sum_{k=1}^K\log \Gamma(\psi_k)-\log\Gamma\left(\sum_{k=1}^K\psi_k\right)\\-\left(K-\sum_{k=1}^K\psi_k\right)\boldsymbol{\Psi}\left(\sum_{k=1}^K\psi_k\right)-\sum_{k=1}^K\left(\psi_k-1\right)\boldsymbol{\Psi}(\psi_k)
+\sum_{r=1}^R\log \Gamma(\varphi_r)-\log\Gamma\left(\sum_{r=1}^R\varphi_r\right)\\-\left(R-\sum_{r=1}^R\varphi_r\right)\boldsymbol{\Psi}\left(\sum_{r=1}^R\varphi_r\right)-\sum_{r=1}^R\left(\varphi_r-1\right)\boldsymbol{\Psi}(\varphi_r)
+a_\alpha-\log b_\alpha+\log \Gamma(a_\alpha)+(1-a_\alpha)\boldsymbol{\Psi}(a_\alpha)\\
+a_\beta-\log b_\beta+\log \Gamma(a_\beta)+(1-a_\beta)\boldsymbol{\Psi}(a_\beta)
\end{multline*}
}
which is used to monitor for convergence of the algorithm.

\end{document}